\documentclass[12pt,a4paper,twoside,%
BCOR12mm,DIV12,%
headsepline,%
cleardoubleempty,chapterprefix,parskip,%
liststotoc,idxtotoc,bibtotoc]{scrbook}

\AfterPackage!{hyperref}{\PassOptionsToPackage{plainpages=false,pdfpagelabels,breaklinks}{hyperref}}
\AfterPackage!{hyperref}{\usepackage{breakurl}}
\usepackage{url}
\makeatletter
\g@addto@macro{\UrlBreaks}{\UrlOrds}
\makeatother
\usepackage{cite}

\usepackage[ngerman, english]{babel}
\usepackage[utf8]{inputenc}
\usepackage[T1]{fontenc}
\usepackage{mathptmx}
\usepackage[scaled=0.92]{helvet}
\usepackage{courier}
\usepackage{amsmath}

\usepackage{amsfonts}
\usepackage{amstext}
\usepackage{subfigure}
\usepackage{color}
\usepackage{graphicx}
\usepackage{ifpdf}
\usepackage{makeidx}
\usepackage{nomencl}
\usepackage{setspace}

\usepackage{lscape}
\usepackage{tabularx}
\usepackage{booktabs}
\usepackage{float}      % Ability to make tables inline
\restylefloat{table}    %
\usepackage{colortbl}   % Make rows colorful

\usepackage{listings}
\usepackage{xcolor, colortbl}
\definecolor{green}{rgb} {0.23, 0.67, 0.2}
\definecolor{orange}{rgb}{0.95, 0.57, 0.0}
\definecolor{red}{rgb}   {0.7, 0.1, 0.1}
\definecolor{white}{rgb} {1, 1, 1}
\newcommand{\colorspacing}{\qquad \qquad}

\newcommand{\green}{\cellcolor{green}\colorspacing}
\newcommand{\orange}{\cellcolor{orange}\colorspacing}
\newcommand{\red}{\cellcolor{red}\colorspacing}
\newcommand{\unknown}{\cellcolor{white}\colorspacing}

\makeatletter
\renewcommand*{\ext@figure}{lot}
\let\c@figure\c@table
\let\ftype@figure\ftype@table
\let\listoftableandfigures\listoftables
\makeatother

\addto\captionsenglish{ % Change heading
    
}
\newcolumntype{Y}{>{\centering\arraybackslash}X}

\ifpdf
    \definecolor{brown}{cmyk}{0, 0.81, 1, 0.60}
    \hypersetup{%
      pdftitle={Design of Distributed Voting Systems}, %%%% HIER AENDERN!!!
      pdfsubject={Master Thesis},
      pdfauthor={Christian Meter},           %%%% HIER AENDERN!!!
      pdfkeywords={electronic voting, construction, design},      %%%% HIER AENDERN!!!
      colorlinks=false, urlcolor=blue, citecolor=brown,
      bookmarksnumbered=true,
    }
\else   
\fi

\begin{document}
\onehalfspacing

\frontmatter

\pdfbookmark[0]{Title}{tit}
% Titelseite

\begin{titlepage}
  \centering
  \includegraphics[width=5cm]{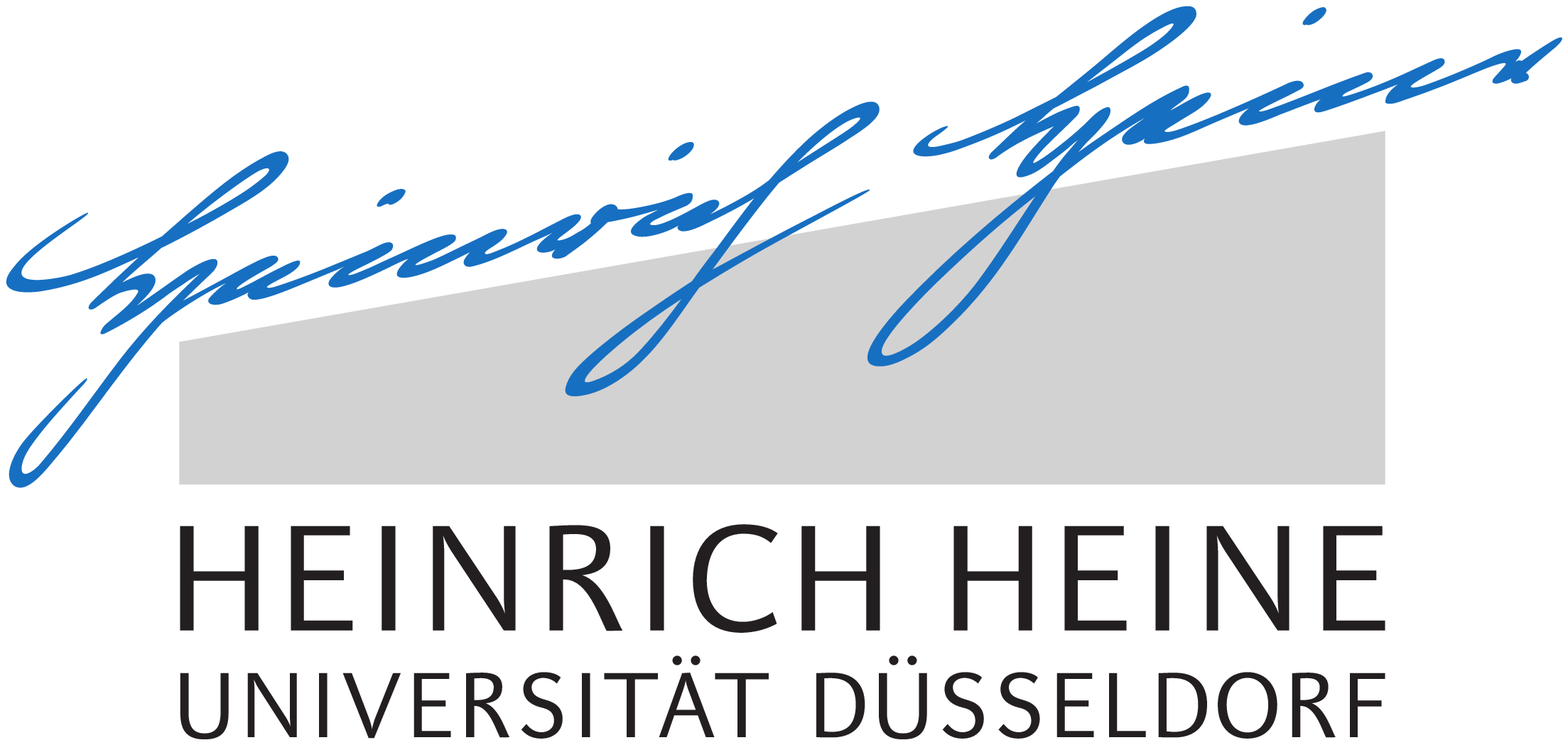}\\

  \vfill
  \huge
  Design of Distributed Voting Systems\\*[40pt]
  \normalsize

  \vfill
  \large
  Masterarbeit\\[0.25em]
  \normalsize
  von\\
  \Large
  Christian Meter\\

  \vspace{5mm}
  \normalsize
  aus\\ Remscheid\\[1cm]
  vorgelegt am\\[5mm]
  Lehrstuhl für Rechnernetze und Kommunikationssysteme\\
  Prof.\ Dr.\ Martin Mauve\\ 
  Heinrich-Heine-Universität Düsseldorf\\[0.5cm]
  24. September 2015\\[0.5cm]
  Betreuer:\\
  Philipp Hagemeister, M.\,Sc.
    
\end{titlepage}

%%% Local Variables: 
%%% mode: latex
%%% TeX-master: "master"
%%% End: 

\cleardoublepage

\pdfbookmark[0]{Abstract}{abstract}
\begin{center}
\huge Abstract
\end{center}

Countries like Estonia, Norway or Australia developed electronic voting systems, which could be used to realize
parliamentary elections with the help of personal computers and the Internet. These systems are completely different in
their design and their way to solve the same problem.\\
In this thesis, we analyze some of the largest real-world systems, describe their building blocks and their general
design to focus on possible problems in these electronic voting systems.

Furthermore, we present a template for an e-voting system, which we designed to try to fulfill the preliminaries and
requirements of a secure electronic voting system. We use the experiences and the building blocks of existing systems to
combine them to another more secure system. Afterwards, we compare our concept with real-world systems to evaluate the
fulfillments of the requirements. Conclusively, we discuss the occurring problems when designing a secure system.

Peer-to-peer networks provide many advantages, like decentralization, which might be applicable to electronic voting
systems. Therefore, we take a look on the distributed database called \emph{blockchain}\footnote{The blockchain is a
distributed database, which was first introduced with the Bitcoin protocol.} and the usage in a peer-to-peer voting
system.\\
Our contribution to this topic is a modification of the proof-of-stake, which enables the usage of common devices, like
smartphones or tablets, for the blockchain verification and inclusion of new ballots to the chain. This proof does not
need much computing power and has a lower carbon footprint than the proof-of-work in the Bitcoin protocol.

\cleardoublepage

\pdfbookmark[0]{Acknowledgements}{ack}
\begin{center}
    \huge Acknowledgments
\end{center}
A lot of people supported me during my work on this thesis to whom I wish to express my gratitude.\\
Thanks to all my friends, who volunteered to proof-read this thesis, namely Frank Heisig, Philip Baues and Alexander
Schneider. It must have been very hard work, since my English is not the best around, but you kept fighting through the
pages.

I also thank Alex for the discussions, fighting and arguing in our office, which led to a deeper understanding of this
topic.

Special thanks to Philipp Hagemeister, who advised this thesis and always supported me. Also thanks to Prof. Martin
Mauve, who made this thesis possible.

\cleardoublepage

\pdfbookmark[0]{\contentsname}{content}

\tableofcontents

\listoftableandfigures

\mainmatter

\cleardoublepage

%%%%%%%%%%%%%%%%%%%%%%%%%%%%%%%%%%%%%%%%%%%%%%
%%    Beginning of the main document        %%
%%                                          %%
%%    Include your tex-files with \input{}  %%
%%%%%%%%%%%%%%%%%%%%%%%%%%%%%%%%%%%%%%%%%%%%%%

\chapter{Motivation}

\section{Traditional Voting}
One basic principle of a democracy is an equal and fair voting system: eligible voters are allowed to freely vote for
their favorable party or candidate. This is one of the pillars of our political system and therefore needs to be
guaranteed in a democracy.

\subsubsection{Some Problems in Parliamentary Elections}
\label{problems-parliamentary-elections}
Democracy and voting are great ideas, but the classical paper ballots are prone to fraud; ballots can be counted
incorrectly or ballots sent via mail might get lost in transit. To show examples for failure or fraud, we focus for this
list on the parliamentary elections from Germany in 2005, because they are well documented. Examples are taken
from~\cite{ZeitWahlmanipulation2013}:

\begin{itemize}
    \item First counting in Bochum-Langendreer marked 491 of 689 votes as invalid. Two minor parties were announced as
        the strongest in this district. After recounting the ballots, only 13 ballots were marked as invalid. A
        different, third party became strongest party.
    \item In one state the ballots were not correctly assigned to the parties, which led to a bad result for a small
        party.
    \item Paper ballots sent via postal mail take a long time until they are tallied. Observations showed that even ten
        days might not be sufficient to request and send the ballot back before the election ends.
    \item An external company was delegated to distribute paper ballots for one city. Unfortunately, they sent 50,000
        ballots to the wrong recipients. Due to this error 10,533 ballots became invalid.
\end{itemize}

These are only a few examples for potential problems with traditional paper voting and they are not the only exceptions.
This does not mean that all elections are compromised or completely insecure.

\subsubsection{High Cost}
Another factor are the costs of an election. We focus on the numbers from Germany again.\\
The parliamentary election for the Bundestag in 2009 did cost about 67 million Euros in total. Cities with less than
100,000 eligible voters received 0.48 Euro per voter, bigger cities even 0.74 Euro~\cite{BundWahlkosten2011}.
Additionally, volunteers, who support an election, received another monetary compensation for their help. This is a
massive amount of money being normally spent every parliamentary election.

One possible solution to reduce the costs and to optimize the general voting process is the usage of computers.

\section{Electronic Voting Systems}
As technology evolves, it is obvious to consider about using computers for elections. In this thesis we will focus on
\textbf{distributed voting systems}, which we define as \textbf{systems using the Internet to realize political
elections}. To access these voting systems, each eligible voter can \textbf{use her own device}, for example personal
computer, smartphone or tablet. These systems will also secure and anonymize the ballots to ensure the election, which
fulfills the democratic rights of each citizen. We take a further look at these requirements and constructions in the
next chapters.

Electronic voting systems attempt to be as easy to use and secure as ideal traditional elections and attempt to
eliminate the human errors described in~\ref{problems-parliamentary-elections}. This is hard to achieve, because
electronic voting systems need a strong encryption to guarantee security, integrity and anonymity of the vote. This must
be ensured and still result in a user-friendly application, which is often hard to achieve.\\
But to assume that traditional elections are completely secure and correct is also questionable, as we already showed in
section~\ref{problems-parliamentary-elections}. So, this is a good opportunity to think about reinventing elections with
the help of computers and cryptography.

One of the main advantages of electronic voting systems is the chance to call a completely verifiable election, which
means that all voters are able to verify if their vote was properly counted and even that the complete election was
properly tallied.\\
Some countries use dedicated voting machines, which are used to place votes in polling stations. These voting machines
are exclusively used for the voting process and can either tally the votes electronically or create strips of papers
with the voter's choice, which must later be tallied. Usually, it is not possible to verify tallying steps of these
black boxes, because the companies do not provide details about the implementation of their machines; only the main
developers have access to the source code and know in detail, how these machines operate.\\
After an analysis of 74 voting machines, the Chaos Computer Club (CCC), which is Europe's largest association of
hackers, summarized their results with one short quote \cite{CCC2006}:

\begin{quote}
    \emph{``Trust is a good thing, control not possible'' (CCC, 2006)}
\end{quote}

The CCC observed in 2006 a pilot project in Cottbus, Germany, where voting machines were used. They explained in their
analysis of this election that with these issues in security and verifiability, voting machines should be banned and
not be used in any election.\\
Missing verifiability led to the prohibition of current voting machines for elections in Germany. As long as the
essential steps of the voting process are not in public verifiable by a typical citizen, voting machines are forbidden
in parliamentary elections~\cite{Bundeswahlleiter2009}. These are also the reasons why we do not consider voting
machines in this thesis.

\subsubsection{Electronic Voting Systems in the Real World}
Some governments already implemented electronic voting systems and use them for parliamentary elections. For example
Estonia has several years of experience in this field and successfully uses electronic voting for all of their
elections. Other projects encountered, but they all had big security issues and were often cancelled. That the Estonian
electronic voting system is still being used in practice does not mean that this voting system is secure. We will
analyze it in section~\ref{estonia}.

We feel confident that many countries will use electronic voting systems in the future to realize their elections,
because this technology could heavily improve the voting process. Therefore, it is essential to analyze existing
systems, learn from their experiences and try to solve the issues which emerged during their trials, which is the core
of this thesis. We also describe basic approaches to realize a voting system with clients and servers and give a brief
view into a peer-to-peer approach using the blockchain.

\section{Structure}
In chapter~\ref{preliminaries} we define the preliminaries and requirements of an election. This also includes some
assumptions we had to include to realize a voting system. Since security, anonymity and integrity must be guaranteed by
computers, we have to use cryptography to solve these issues. The cryptographic primitives used by many voting systems
are described in chapter~\ref{primitives}.\\
Chapter~\ref{systems} contains a selection of popular e-voting systems, a description of their design and their major
problems. These systems are compared with each other to provide a brief overview of their building blocks.

With the knowledge of these real-world systems, we choose building blocks for a secure voting system in
chapter~\ref{construction}.\\
During our research, we found a promising approach using the blockchain. We designed a suitable proof-of-work
replacement and described it in the same chapter.

Our evaluation in chapter~\ref{evaluation} analyzes if our construction fulfills the preliminaries and compares our
system with the real-world systems from chapter~\ref{systems}.

In chapter~\ref{conclusion} we summarize our findings and give an overview about future work.

\chapter{Preliminaries}
\label{preliminaries}
Electronic voting systems claim to be at least as secure as ideal traditional voting systems like paper ballots. In
fact, paper ballots (or even special voting machines) have many potential security issues as seen
in~\ref{problems-parliamentary-elections}. With the correct use of cryptography these issues can be limited, which is a
great advantage of e-voting systems. Some requirements have to be fulfilled to make a voting system applicable for the
real-world. This list is based on~\cite{Clarkson2008,delaune2010verifying,kremer2010election} and the systems we
describe in chapter~\ref{systems}.

\paragraph{Availability}
\label{availability}
An e-voting system must remain available during the whole election and must serve voters connecting from their
devices.\\
Especially, the e-voting system must be prepared for high workload, because there will be periods where many voters will
place their vote simultaneously.

\paragraph{Eligibility}
\label{eligibility}
Only eligible voters are allowed to cast a ballot, whilst only one vote per voter counts. If it is allowed to vote
multiple times (also called re-vote), the most recent ballot will be tallied and all others must be discarded.

\paragraph{Integrity}
\label{integrity}
The integrity of the vote must be guaranteed.\\
Voting systems must ensure that the ballots are not altered during any step of the election. Otherwise we can not trust
this system.

\paragraph{Anonymity and Election Secrecy}
\label{anonymity}
The connection between the vote of a user and the user herself must not be reconstructable without her help.

\paragraph{Fairness}
\label{fairness}
Voting systems must ensure that no (partial) results are published before the tallying has ended. Otherwise voters can
be influenced by these results and vote differently.

\paragraph{Correctness}
\label{correctness}
The election results must be properly counted and correctly published.

\paragraph{Robustness}
\label{robustness}
The system should be able to tolerate (some) faulty votes.\\
Attackers might try to cast malicious ballots, but these ballots must be detected. A voting system has to recognize
these ballots to prevent vote-manipulation or attacks on the servers.

\paragraph{Universal Verifiability}
\label{universal-verifiability}
After the tallying process, the results are published and must be verifiable by everybody.\\
The electronic voting system must provide mechanisms to verify the election's outcome. This depends on the building
blocks the system is built upon and must not break other preliminaries.

\paragraph{Voter Verifiability}
\label{voter-verifiable}
The voter herself must be able to verify that her ballot arrived in the ballot box.\\
This ensures that the voter is sure her vote was counted and was not modified.

\paragraph{Coercion Freeness}
\label{coercion-freeness}
\label{receipt-freeness}
Voting systems must provide security mechanisms to prevent a coercer from being able to force the voter to place a vote
for a specific party, candidate etc.\@ or even to see \emph{that} she voted~\cite{Okamoto1998}. This is also called
\emph{receipt-freeness}.\\
A voting system must be built coercion-resistant to guarantee that a voter can place her vote as intended even in the
presence of a coercer. Even vote-selling must be unattractive or too expensive. Coercion is a major problem in voting
systems and we discuss it in detail in subsection~\ref{coercion-open-issue}.

\subsubsection{Summary}
\label{preliminary-coercion-summary}
These requirements are necessary for a secure e-voting system, which adds complexity and makes secure design and a
usable interface more difficult. The big challenge for voting systems is to fulfill as many requirements as possible and
create a secure voting system that is easy enough for everybody to understand and to use.

Coercion and receipt-freeness are the most challenging requirements. On the one hand it is necessary to provide the
option to verify her own vote, but this is always coupled to some kind of receipt. On the other hand a voter must not be
able to prove her choice to a coercer. We will discuss this later in subsections~\ref{discussing-coercion-issue}
and~\ref{eval-voter-verifiability}.

Voter- and universal-verifiability are needed to achieve \emph{end-to-end verifiability}, which is the possibility to
verify the complete voting process. This includes all steps from the composition of the own ballot over sending the
vote to the ballot boxes through the anonymization servers to the tallying process~\cite{Benaloh2015}. It is sufficient
to provide proofs for the separate steps showing that the servers worked as expected (see \emph{zero-knowledge-proofs},
section~\ref{zerknopro}).

\section{Assumptions}
\label{assumptions}
We have to make few assumptions, which are required to make our constructed electronic voting system described in
chapter~\ref{construction} possible and useful. Many systems from chapter~\ref{systems} make similar assumptions
(see~\cite{Demirel2012,Clarkson2008}), which is why we already want to introduce them:

\subsubsection{Assumption 1: The voter's computer can be trusted}
We assume that it is possible to securely run the voting application on the voter's device. This excludes malicious
software, which might be installed on the voter's device and might unobtrusively alter her ballot.

\subsubsection{Assumption 2: The election is correctly set up}
The election must be set up correctly, which means that the candidates and parties are included in the election, there
are only eligible voters allowed to place a ballot and nothing is compromised prior the election. Without this
assumption, the election itself is already non-trustworthy and can not produce a reliable outcome.

\subsubsection{Assumption 3: Not all trustees of the election are compromised}
We describe the election's building blocks in chapter~\ref{construction} and describe how many trustees must not be
malicious for the system to work properly, e.g.\@ it takes at least one trustworthy server in the mix-net to provide
anonymity of the ballots (see~\ref{anonymization-of-ballots}). This assumption shows that a minimum number of the
trustees is trustworthy and this makes a reliable election possible.

\subsubsection{Assumption 4: At least one person verifies the results}
There should be at least one person who verifies the results at the end of an election. This makes it unlikely that the
election has been compromised when at least one person is able to reproduce the result. Since the election's outcome
should be public and verifiable, it does not matter who this person is, but she should publish her results to approve
the outcome or that she found irregularities in the tallying process.

\chapter{Cryptographic Primitives}
\label{primitives}
This chapter briefly describes some of the cryptographic primitives which are used in many electronic voting systems.
These are the building blocks of some of the biggest real-world systems and are used in several different combinations.

\section{Public Key Cryptography}
\label{pubkeycry}
In real world voting systems, the asymmetric cryptography is heavily used to de-/encrypt or sign a ballot. Based on
algorithms like RSA, the ``classical'' way is used to gain advantage of this technique. Thereby, each voter and the
election server maintains a key-pair.

\begin{figure*}
    \includegraphics[width=\textwidth]{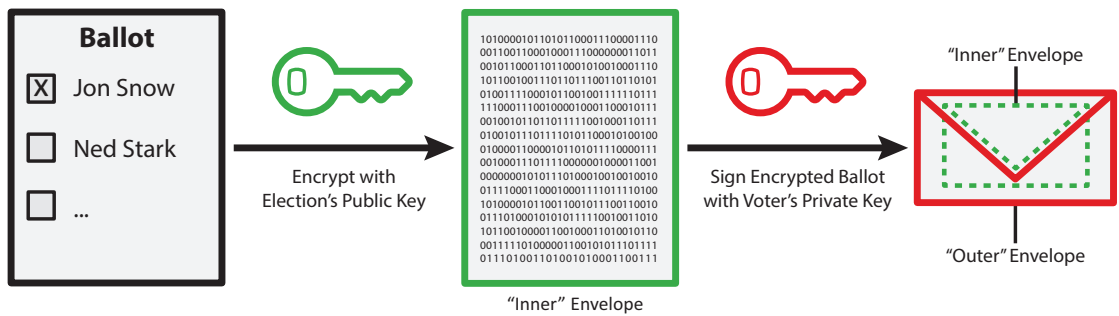}
    \caption{Double Envelope -- a Signed and Encrypted Ballot}
    \label{fig:doubleenvelope}
\end{figure*}

Mostly, the technique of a double envelope is chosen in electronic voting systems, which is being used for postal
ballots: In the inner envelope is the ballot of the voter $m$, which is encrypted with the election's public key
$enc(m)_{pub\_el}$, i.e.\@ it is ``packed'' into a ciphertext. The outer envelope contains the signature of the voter,
who signs just that encrypted ballot with her private key $sig(enc(m)_{pub\_el})_{priv\_voter}$. This is illustrated in
figure~\ref{fig:doubleenvelope}. With this packed ballot, the voter can contact a voting server, which can verify if she
is eligible to vote by checking the signature. If she is eligible, the ballot is stored in the election's database.
Before tallying the ballots, the signature is stripped off and should be passed through a mix-net (see~\ref{mixtyp}) or
similar to guarantee anonymity during and after the tallying process.

\paragraph{Advantages}
The concept of public key cryptography is well understood and generally easy to implement. Therefore, explaining it to
voters is not difficult and there are several libraries existing to be used in the source code of the voting system.

\paragraph{Drawbacks}
A Public Key Infrastructure (PKI) is needed to maintain all public keys of the voters. This can be combined with the
registration for the election and with the validation of the voter's eligibility.\\
It takes a lot of computational power to decrypt all votes, so publishing the results might take a while.

\paragraph{Usage}
Most electronic voting systems rely on public key cryptography (see~\ref{estonia}, \ref{norwayvoting}, \ref{australia},
\ref{civitas}). In general, this is currently best practice as long as it is well implemented and the keys are long
enough. But intelligence agencies, like the NSA, take deep interests in manipulating the RSA standard and bribed the
developers \$10 million to make a manipulated random number generator, based on the RSA's Dual Elliptic Curve, the new
default system~\cite{ReutersRSAManipulation2014}.\\
Therefore, developers must be very careful while implementing RSA in their voting systems and must choose (currently)
safe algorithms for random number generation.

\section{Zero-Knowledge-Proofs}
\label{zerknopro}
Zero-knowledge-proofs (ZKP) are used when Alice wants to prove to Bob that a specific statement is true without
revealing any information, except that this statement is indeed true. Therefore, no knowledge is transferred even if at
least one of them is malicious.\\
This proof can be applied multiple times, while with each execution of it the probability that Alice just pretends to
know the secret significantly decreases and Bob can verify the correctness~\cite{HelgerLipmaa1997, Brandt2005}. It also
decreases the probability that Alice just guessed the correct solution.

ZKPs can be interactive or non-interactive. In the non-interactive variation only one party is actively needed to verify
the proof, whilst in the interactive one both parties communicate together in a certain way.  Voting systems mostly use
non-interactive ZKPs since the voter can verify several steps without needing an active part of the voting system. This
is favorable for the voting system, because it does not need to spend any resources for these proofs, except the initial
resources needed to create the proof.\\
Existing heuristics allow it to transform an interactive zero-knowledge-proof into a non-interactive
ZKP~\cite{bernhard2012not}. These heuristics were exemplary applied to the Helios voting system (see~\ref{helios}).

In the context of electronic voting systems, zero-knowledge-proofs are mostly used to provide verifiability for a
step inside the voting system, e.g.\@ when the ballots are passed through a mix-net server. This is useful, since it
proves the correctness of each task from the anonymization over the tallying process up to the correct decryption for
calculating the results.

\paragraph{Advantages}
Zero-knowledge-proofs Provide the possibility to validate the ballots and enables end-to-end verifiability, when the
proofs are publicly available.

\paragraph{Drawbacks}
The communication in interactive ZKPs contains many messages between voting server and verifier, which leads to a big
overhead for just verifying the proof. But as said it is possible to use non-interactive proofs, which are sufficient
for our purposes.

\paragraph{Usage}
Zero-knowledge-proofs are needed for verification. Therefore, many systems use and combine them with other primitives,
because it is an easy way to verify the results of an operation (for example see~\ref{norwayvoting}, \ref{civitas}).
This primitive is an essential building block to achieve end-to-end verifiability.

\section{Homomorphic Encryption}
\label{homenc}
The homomorphic encryption scheme is a subset of the public key cryptography (see~\ref{pubkeycry}), where
mathematical operations directly on the ciphertexts are possible. These mathematical operations might be the
re-encryption of the ciphertext without changing the content (see~\ref{reencryption}) or the possibility to aggregate
the ciphertexts to add up the values of encrypted votes.\\
For example: assume the ballots $a$ and $b$ are encrypted with a homomorphic scheme to $a = enc(2)$ and $b = enc(3)$.
Than they can be aggregated to $a \odot b$ and this operation leads to the same result as $enc(2 + 3) = dec(a \odot b) =
5$~\cite{Hirt2000}. So, there is no need to decrypt each ciphertext to tally them. The next section describes the
structure of the ballots before we can apply this directly on electronic voting systems.

\subsection{Restrictions and Usage in Electronic Voting Systems}
When using an encryption scheme which uses homomorphic properties allowing the addition on the ciphertexts, like
ElGamal, there is a restriction in the structure of the ballot: the ballots must be encoded with bits before they are
encrypted. This means that the candidate the voter wants to vote for gets a $1$ whilst all other candidates have a $0$
stored in the corresponding position. For example, Alice wants to vote for the second candidate on the list. Her ballot
must look like $010\dots$, which is then encrypted. Vectors can also be used for this data structure, which support more
votes for each candidate.\\
Because of this structure, it is only suitable for elections where \emph{yes} or \emph{no} are possible answers for the
candidates. Write-in-ballots, as they are used in the United States, are therefore not supported. We can not encode a
string, e.g.\@ the name of a candidate, into one bit and the homomorphic addition does not support addition of strings.

Applied to electronic voting system we consider the following example. We have three candidates, Alice wants to vote for
candidate 2, Bob for candidate 3 and Charly also for the second candidate. The resulting ballots are encrypted with the
election's public key:
\begin{align}
    Alice: && a &= encrypt(010) &&\\
    Bob:   && b &= encrypt(001) &&\\
    Charly:&& c &= encrypt(010) &&
\end{align}
When the election ends and the vote count starts, we can easily use the addition on the ciphertexts, which directly
produces the correct outcome of the election. The result must be decrypted with the election's private key and might
look like this:
\begin{align}
    res          &= a \odot b \odot c\\
    decrypt(res) &= 021
\end{align}
This result can be decoded and leads to the expected result. Candidate 1 has zero votes, candidate 2 has two and only
one person voted for candidate 3. If we would allow write-ins in this example, it would not be possible to simply add
the ciphertexts, because we can not apply the simple addition on strings. Assuming Alice is candidate 2 in this example,
the homomorphic property can not aggregate the two ballots containing ``Alice'' + ``Alice''. In an election, we would
expect the sum ``2'', but this is not possible with this property and this is also the reason why write-ins are not
supported.

False inputs might cause unexpected errors, since the falsy composition is not compatible with the homomorphic addition.
Therefore, most voting systems use Zero-Knowledge-Proofs (described in~\ref{zerknopro}) to guarantee that they encrypted
a correct ballot matching the chosen data structure.

\subsection{Re-Encryption of Ciphertexts}
\label{reencryption}
Some encryption schemes enable re-encryption, again for example ElGamal. This is another mathematical component,
additional to e.g.\@ \emph{encrypt()} or \emph{decrypt()}, which re-randomizes the random factor in a
ciphertext~\cite{Golle2004}. This results in a different-looking ciphertext, although it still contains the same
content.\\
For example: a ciphertext $\{m\}_k^r$ encrypted with the public key $k$ and a random factor $r$ could be re-encrypted
with the same $m, r, k$ and a new random factor $r'$ to $\{m\}_k^{r+r'} = renc(m, r')$~\cite{Grewal2013}. This operation
does not need the private key of the election and is therefore not decrypted, thanks to the homomorphic property for
example provided by ElGamal~\cite{Golle2004}. This is a basic primitive for mix-nets (explained in the next
section~\ref{mixtyp}), because a mix-net takes the ballots, creates a permutation of them and re-encrypts them for
anonymization. Therefore, we can achieve anonymity of a set of ballots with this property when we rearrange the order of
the ballots and change the outward appearance of the ciphertexts. To guarantee that the re-encryption is correct and the
ballot's integrity is ensured, the re-encryption can be made verifiable with the help of zero-knowledge-proofs
(see~\ref{zerknopro}).

\paragraph{Advantages}
These encryption schemes with homomorphic properties have several benefits for electronic voting systems, which depend
on the algorithms. Being able to aggregate the ciphertexts simplifies the tallying process, since only one decryption is
needed after all ballots were aggregated.\\
Different schemes of homomorphic encryption also enable a number of mechanisms, like secret-sharing or the re-encryption
of the ballots, which is heavily used in voting systems which use mix-nets for anonymization.\\
Keys generated with a homomorphic scheme, can also be used normally as seen in the public key
cryptography~\ref{pubkeycry}.

\paragraph{Drawbacks}
Some schemes are only suitable for elections where \emph{yes} or \emph{no} are possible answers.\\
Another big drawback is the computing time needed to aggregate homomorphic encrypted ballots. This is very complex and
might not be applicable on big amounts of encrypted ballots. Kristian Gj{\o}steen from the Norwegian University of
Science and Technology is researching for the Norwegian voting system to massively reduce the size of the ciphertexts to
decrease the computational time and presents some mathematical approaches~\cite{Gjosteen2013}.\\
The developers of Civitas tried to benchmark the effort needed to decrypt the ballots with respect to different
parameters~\cite{Clarkson2008}.

\paragraph{Usage}
Homomorphic encryption is heavily used in the Norwegian e-voting system (see~\ref{norwayvoting}), where the homomorphic
property is used to count the ballots. They have the problem that the ciphertexts are too big and decryption takes too
much time.\\
Civitas (see~\ref{civitas}) uses ElGamal for their re-encryption scheme after the votes passed the mix-net
(see~\ref{mixtyp}).

\section{Mix-Nets}
\label{mixtyp}
Mix-net based voting schemes use the same technique as Tor to anonymize a user's traffic while surfing the Web: Multiple
mix-servers are used to remove connections to the voter. They shuffle and re-encrypt the ballots to make them look
different than they looked the step before. The correctness of the result can be verified using zero-knowledge-proofs,
which each authority has to publish after shuffling. The mix-servers can be used to anonymize the ballots, because these
servers remove the connection of the voter's signature and her vote and re-encrypt the ballots.  As long as there exists
at least one honest mix-server, the anonymity of the voter is guaranteed~\cite{Sako1995, Angels2014}.\\
Some voting systems use this technique as an extension to achieve anonymity~\cite{Neff2001}.\\
Re-encryption is needed, because otherwise the rearranged ballots will look the same, but in a different order.
Following the notation from subsection~\ref{reencryption}, the sequence of encrypted ballots $S = m_1, \dots, m_n$ are
formed to a different sequence $S' = m'_1, \dots, m'_n$, which is a re-encryption of $S$. Also, the order of the ballots
has changed with the permutation $\sigma$ of $\{1, \dots, n\}$. The new random factors $r'_1, \dots, r'_n$ are then used
to re-encrypt $S$ to get $S'$: $m'_1 = renc(m_{\sigma(1)}, r'_1), \dots, m'_n = renc(m_{\sigma(n)},
r'_n)$~\cite{Grewal2013}. As a result, $S'$ is returned, which can later be decrypted with the election's private
key~\cite{Golle2004}. There is no private key needed to re-encrypt the ballots.\\
All these steps can be verified with ZKPs, which each server in the mix-net has to publish.\\
Mix-nets require an encryption scheme, which supports re-encryption, like ElGamal.

\paragraph{Advantages}
Mix-net servers provide anonymity with a simple, well-known procedure and are robust against attacks on the voter's
identity. These servers can easily be distributed among multiple and independent authorities. As long as one of these
authorities is honest, the mix is successful and the connection between the voter and her ballot is removed. As a result
all ballots are anonymized.

\paragraph{Drawbacks}
Ideally, we need many dedicated servers for a mix-net to perform the mixes and to calculate the ZKPs. Also the
cryptographic operations need many resources, which are not deniable (see~\ref{reencryption}).

\paragraph{Usage}
Mix-nets are mostly used when the ballots are encrypted with a double-envelope scheme (like in~\ref{norwayvoting},
\ref{civitas}), where the voting system wants to anonymize the ballots before tallying (and publishing). Then the
signature is stripped off and the mix-nets guarantee that it is no longer possible to reconstruct the connection
between the voter and her ballot.

\section{Secret Sharing and Threshold Encryption}
\label{diselg}
To achieve distributed trust, the election's private key can be distributed among a specific number of trustees.
Therefore, to decrypt the ballots, there is a specific threshold of trustees needed. For example: as long as $n$ out of
$m$ authorities are not corrupt, the keys can be restored and used for the tallying process~\cite{Brandt2005,
Frankel1998}.

\paragraph{Advantages}
Distributing the key-pairs leads to a more secure and confidence inspiring voting system, because to break the election,
$n$ ballot-tallying trustees must be corrupt, which is much more difficult for an attacker than just compromising a
single trustee.

\paragraph{Drawbacks}
If $m - n + 1$ trustees are compromised or simply refuse to cooperate with the other trustees, the secret is lost and
can not be created. Systems, like the Norwegian e-voting system (see~\ref{norwayvoting}), set $n = m$ which means that
all trustees have to cooperate. In this case it is sufficient that exactly one non-cooperating trustee could lead to a
non-reconstructible secret, like the private key.

\paragraph{Usage}
The Estonian electronic voting system (see~\ref{estonia}) already implements it to create the private key. All
trustees $n = m$ are needed to create the private key. This is the most secure option when distributing parts of the
key, because this is the highest possible value for the threshold and no subset of them are able to create the key.

\section{Everlasting Privacy}
\label{everlastingprimitive}
A critical question in encryption is what happens to the privacy when the algorithms used for encryption are no longer
secure and the ballots can than be decrypted without the secret key. This might be possible when computing power
increases or brute force attacks allow the decryption of the ballots without the key. Research in the field of
everlasting privacy focuses on this topic to keep the ballot's content private~\cite{Arapinis2013}.

This is useful in several cases: Firstly, even when the ballots are published in the end of an election, nobody would
ever be able to decrypt it without the private key. Just think about a new government, which wants to sentence someone
for his ballot, which was placed many years in the past. It is therefore very important to keep the ballot's content
secret. Secondly, one can think about an attacker, who compromised one part of the system, where the ballots pass by,
e.g.\@ the firewall. This attacker might copy and store the bypassing ballots to decrypt them in the future, when there
is enough computational power available or the encryption algorithm is proven insecure.\\
In both cases the anonymity can be lifted, even some years in the future.

Everlasting privacy must directly be used for composing the ballots, before the ballot is sent to the voting system.

\paragraph{Advantages}
The ballot's content is kept secret through the complete voting process and is only decryptable with the election's
private key. But most important is that the ballots are also secure against attacks and vulnerabilities in the near
future.

\paragraph{Drawbacks}
The cryptography behind everlasting privacy is hard to understand, because it mostly uses the applied pi
calculus~\cite{Arapinis2013}. Also, we found no libraries for popular programming languages supporting the usage of
everlasting privacy, which makes it difficult for the developers to use this primitive without having a deep
understanding of cryptography.

\paragraph{Usage}
Some scientists faced the problem and developed additions to existing systems~\cite{Demirel2012,
EverlastingPretAVoter2013} or even developed a complete voting scheme using everlasting privacy~\cite{Privacy2006,
Demirel2012a}. But the voting systems which have been already used for real-world elections (see
chapter~\ref{systems}), do not use everlasting privacy at all.

\section{Blind Signatures}
\label{blisig}
In a system using blind signatures, a correctly composed ballot is signed by an authentication server without needing to
decrypt it.\\
In the first steps the voter prepares her vote, adds a blinding factor to it and authenticates at an authentication
server of the voting system. This server checks if the voter is allowed to vote, has not voted before and correctly
composed her ballot. If that is true, the authentication server signs the encrypted vote~\cite{Angels2014}. After this
step the voter can remove the blinding and has the correctly signed and well-formed ballot.

To prove well-formedness of a ballot, the voter has to add a zero-knowledge-proof to her blinded vote
(see~\ref{zerknopro}). This proof ensures that she correctly composed her ballot and correctly added the blinding
factor. The authentication server needs to verify the proof and then signs it. This step is necessary, because only
well-formed ballots contain the designated input (e.g.\@ exactly one vote for a valid candidate) and can later be
tallied.\\
After the voter receives her blinded and signed vote, she can subtract the random factor out and has her valid vote
prepared for tallying.

\paragraph{Advantages}
Blind Signatures are very simple and easy to understand, because they can be applied on normal scenarios with offline
letters: Alice prepares her vote on a special letter, folds and seals it and wants Bob to sign it. Bob sees that the
correct letter was used and without breaking the seal he signs the letter and sends it back to Alice. She now has her
properly sealed ballot with the signature of Bob and she is now able to send the vote to the tallying station. The
station verifies the signature and counts the vote.

\paragraph{Drawbacks}
Most voting systems allow duplicate voting to prevent coercion. But blindly signed ballots have no connection to the
original voter and therefore it is not possible to find other cast votes by the same voter to drop all except the last
of her ballots. This is why this primitive is only used in some theoretical schemes, but not in real-world voting
systems.

\paragraph{Usage}
None of our analyzed voting systems uses blind signatures, because they all allow re-voting to override old ballots.

\chapter{Systems}
\label{systems}
This chapter describes several real-world e-voting systems, which are used or were supposed to be used for
parliamentary elections in the last years.

In the end of this chapter, we will shortly focus on academical proof-of-concepts, which provide promising ideas in
enhancing current e-voting systems.

\section{Estonian I-Voting System}
\label{estonia}
Estonia is a modern country, which heavily relies on the Internet. Nearly everything is possible with the Internet
combined with their electronic national ID cards (eID). These ID cards are used for the e-voting system. The
government council election of 2005 was the first election where their citizens were able to vote via the
Internet~\cite{Maaten2004}. Estonia still maintains and uses their I-voting system for the parliamentary elections.

\paragraph{ID cards and PKI}
\label{estoniapki}
The ID cards are realized on a Java chip platform, containing a 2048-Bit PIN-protected RSA key-pair and creating
signatures with SHA1/SHA2~\cite{Trub2013}. This conforms to common security practices in the Web and can easily be used
for authentication, encryption, signatures, etc.\\
Since the government distributes the ID cards, they keep track of the public keys used by the citizens. Therefore,
authenticating at the electronic voting system and validating the eligibility is easy, because the voter just has to
create her signature with the ID card, send this signature to the application's authentication servers and is
authenticated through the government's PKI.

\subsection{Application}
\label{estoniaapplication}
The application \emph{I-voting Client} is developed for most popular operating systems including Windows, Linux and Mac
OS X. These applications guide the voter through the voting process. The published version of this system already
includes the election's public key for encryption and the complete communication with the election's data center is
served via a HTTPS connection.\\
Detailed instructions, guidelines, videos\footnote{\url{https://vimeo.com/112041827}} and statistics for the voters can
be found on a special website\footnote{\url{https://www.valimised.ee}}.

The core server code of the Estonian e-voting system is made open source, whilst the I-voting clients, the script to
post a vote and the drivers for the hardware security module (HSM) are kept closed. The HSM is used to decrypt and count
the votes and to output the official results~\cite{Halderman2014}. Therefore, most parts of the application can be
crowd-reviewed for security issues, but without reviewing all parts of the source code, complete trustworthiness cannot
be achieved. A snapshot of the core server code is published on GitHub right before the election
starts~\cite{EstoniaGithubRepo}. The maintainers do not want to publish the I-voting clients, because they are afraid
that this would make it too easy for an attacker to build a fake voting application, which completely looks like the
original one~\cite{Halderman2014}. It is currently unknown why the maintainers do not publish the drivers for the HSM
and the script to post an e-vote.

\subsection{Voting Process}
\label{estoniavoting}
The voter has to download the application via the Internet from one of the authorized websites. As a first step, she
needs to authenticate with her electronic ID or her mobile ID (via smartphone). If she is eligible, she gets a list with
the candidates and can pick one. This vote is being encrypted with the election's public key, signed with the voter's
private key (\emph{double envelope}, see~\ref{pubkeycry}) and sent to the Vote Forwarding Server, which forwards the
correctly encrypted ballot to the Vote Storage Server and leaves a log entry on a special Log Server. These three
servers are deployed in a data center controlled by the election authorities.\\
For verification of the vote, the Voting Client generates an unguessable token packed into a QR Code, which can be
scanned with the Voting App installed on the voter's smartphone. Scanning this code with the voter's smartphone shows
for which candidate she voted for. This is only possible for three times and within 30 minutes after sending the ballot
to the data center and only as long as the eID card is still plugged into the card reader.\\
The voter is allowed to vote multiple times via the I-voting client. This prevents coercion and vote buying as the
coerced vote is invalid after a new ballot has been cast; only the last vote is being tallied. It is also possible to
visit a classic ballot box and vote via paper, which makes all electronic ballots of this voter invalid and uses the
paper ballot instead, because the paper ballot has a higher priority.

\subsection{Tallying Process}
The ballots are composed as double envelopes, therefore the connection between the voter and her vote still exists. As a
next step, this connections must be removed before the ballots are decrypted. So, the voter's signature needs to be
stripped off from the encrypted ballots. These steps are performed on the Vote Storage Server and as the ballots are
anonymized, they are burned to a DVD and transferred to the air-gapped Vote Counting Server. This separate server is
chosen for security reasons, because the isolated Vote Counting Server has no connection to the network, which
drastically reduces the possibility to compromise it or to inject malicious code. Moreover, this server is connected to
the HSM module, which is needed to decrypt the ballots.\\
The election's private key is distributed over multiple authorities as seen in~\ref{homenc}. All of these authorities
have to cooperate to recreate the private key. With this key the ballots can be decrypted and tallied.\\
As a last step, the election's outcome and statistics about the election are published on the official
website~\cite{EstoniaStatistics2015}.

\subsection{Public Evaluation}
In the last parliamentary elections in 2015, 64.2\% (577,910 voters) of the eligible voters participated actively in the
election. 30.5\% (176,491 voters) of these voters used I-voting to place their vote~\cite{EstonianIVotingStatistics}.
This underlines the acceptance of I-voting in the Estonian population.

\subsection{Security Problems}
\label{estonia-security-problems}
The Estonian system uses several cryptographic primitives, but there are many security issues which we will now shortly
describe.

\subsubsection{Operational Security}
Alex Halderman and three members of his team from the University of Michigan were officially accredited observers of an
election in October 2013. They observed the operations in the data centers during the election. This team published a
homepage explaining their results to the citizen~\cite{EstoniaEvoting2015} and a paper showing procedural and
operational security issues~\cite{Halderman2014}:

\begin{itemize}
    \item ``unclean'' computers -- personal computers were used to prepare the election software for the public.
    \item lack of security personnel -- webcams are installed for security, but there was no 24/7 personnel observing
        it.
    \item WiFi passwords are pinned to a wall and recorded by a camera. These cameras even
        recorded the keyboard of a maintainer typing in the root password for one of the servers.
\end{itemize}

Since the developers of the software use their own private computers and download software over an insecure channel, it
might be possible for an attacker to serve manipulated software from an untrusted source. This opens a security issue,
where the attacker might take over the control of the developer's machines with the help of the manipulated software and
distribute the compromised voting application to the voters. Since the application is not completely open source, the
attacker might hide the malicious code in the closed parts of the code.

Another big issue is that administrators are often alone at the servers. The operators of the Estonian system specified
that at least two administrators have to be together in one room while working on the servers. This should reduce the
possibility of a malicious administrator to inject malware, modify the servers or manipulate the votes. The analysis of
the Estonian system proved that the administrators did not comply with these regulations, which makes it potentially
susceptible for insider-attacks (see~\ref{insider-attacks}).

\subsubsection{Technical Security}
This system is vulnerable against \emph{state-level attackers}, like intelligence agencies: These attackers have access
to big parts of the network traffic, enough capacities to store and analyse it and perform timing
attacks~\cite{EnglishErinandHamilton1996}. Therefore, an attacker could analyse the timings of the packets needed for the
communication with the voting servers to prove with a certain percentage that a voter placed her vote. We described this
attack in section~\ref{eval-anonymous-connection}. An attack like this breaks the requirement that a voting system must
guarantee \emph{coercion-freeness} (see chapter~\ref{preliminaries}).

\subsubsection{Centralized Infrastructure}
All servers are concentrated in one data center and the system is therefore vulnerable against DDoS or similar attacks.
Distributing the servers would lead to a higher availability, but will be more expensive and complicated than just
keeping everything central in one center. Also securing distributed servers and their communication is more complicated.

\subsubsection{Client-Side Attacks}
\label{estonia-client-side-attacks}
Like in all applications shipped and executed on private computers of the voters, client-side attacks are possible,
which aim to manipulate the voter's computer. Again, intelligence agencies, like the BND in Germany, have a separate
budget just for buying zero-day-exploits (even from the black market) to gain control over computer
systems~\cite{ZeitBNDExploit}. Therefore, manipulating votes directly on the voter's device is possibly undetected by
the voter.\\
Halderman and his team also describe a \emph{Ghost Attack}, where the compromised computer places a vote shortly before
the 30 minutes passed after the voter was able to verify her vote with the Voting App on her smartphone. This attack
required that the eID card was still plugged into the card reader.

% Summarize Estonian System
\begin{table}
    \centering
    \begin{tabularx}{0.95\textwidth}{l l}
        \toprule
        \emph{Estonia}          & \\
        \midrule
        Authentication          & Electronic ID card\\
        Voting Policy           & Multiple votes\\
        App Structure           & Client/Server with native app\\
        Distributed             & Dedicated servers, but all in one data center\\
        Development Model       & Partly open source\\
        Encryption Scheme       & Public key encryption\\
        Ballot Anonymity        & Signature stripped off from the ballot\\
        Tallying Process        & Separate server with offline-kept private key\\
        Voter Verifiability     & 30 minutes after vote, with smartphone app\\
        Universal Verifiability & No\\
        \bottomrule
    \end{tabularx}
    \caption{Summary of Estonian Voting System}
    \label{table:estonia}
\end{table}

\subsection{Summary}
The ID cards use (currently) secure cryptographic functions with a PKI controlled by the government. This enables a
simple authentication at the voting system for each eligible citizen. No credentials are sent via postal mail or a
similar insecure transportation medium.\\
They used very simple and commonly used cryptographic primitives, so that this voting system ``\emph{can be realised
with the help of IT knowledge existing in Estonia}''~\cite{Ansper2010}, which is a great advantage, because the
government does not need to trust other countries or companies to develop their voting system.\\
Transparency is achieved in a well-documented voting process and with the open source code.\\
For voter verifiability the application displays a QR code after the ballot was sent, which can be scanned with a
mobile phone to display the content of the vote.\\
Allowing multiple votes reduces coercion, however it does not protect against timing attacks and the resulting
possibility to observe a group of voters and prove that they voted, no matter for which party.

% 2015: 176,491 Votes in Estonia

% double envelope as in pubkeycry ✓

% allowed to cast multiple votes ✓
%   coercion ✓
%   vote-buying ✓
% paper ballots at higher priority than e-voting ✓

%% Voting Process
% based on their National ID Cards ✓
% voting applications for many OSs ✓
% public key packed into the applications risky?
% two envelops for ballot ✓
%   inner: ballot, encrypted with elections public key ✓
%   outer: signature of voter (later stripped off) ✓
% strip off signature if voter’s eligibility is established ✓
% leaves a set of anonymous votes ✓
% threshold encryption ✓
% these are burned to a DVD and tallied on on the Vote Counting Server ✓
% send back an unguessable unique token to the voter, with which the voter can verify her vote with a smartphone ✓
% coercion prevention: let the voter vote multiple times, last vote counts ✓

% Statistics for Estonia: http://www.vvk.ee/voting-methods-in-estonia/engindex/statistics/

%%%%%%%%%%%%%%%%%%%%%%%%%%%%%%%%%%%%%%%%%%%%%%%%%%%%%%%%%%%%%%%%%%%%%%%%%%%%%%%%%%%%%%%%%%%%%%%%%%%%%%%%%%%%%%%%%%%%%%%%

\section{D.C.\@ Digital-Vote-by-Mail Service (DVBM)}
\label{dvbm}
Washington, D.C.\@ developed a pilot electronic voting system in 2010. They started a mock election and challenged the
community to test the security of this system. It was again Alex Halderman with Scott Wolschok and his team from the
University of Michigan, who participated in this test and found some critical security issues in the system, whereas the
pilot project was cancelled and not used for the official election~\cite{Wolchok2010}.

\subsection{Application and Voting Process}
DVBM is an open source web application written for Ruby on Rails, using an Apache web server and MySQL as database.
To connect to the web server, an Intrusion Detection System (IDS) and a firewall have to be passed through a secure
HTTPS connection. An Application Server runs the DVBM software and serves a PDF file for the voter, which she needs to
download, mark her candidate and upload the PDF file back to the server. The PDF is being encrypted on the election's
server with the election's public key and passed through another firewall to the Database Server, which stores the
ballot.\\
To get the PDF, the voter has to authenticate herself with the credentials she received prior to the election via postal
mail. These credentials contain a voter ID, registered name, ZIP code and a 16-characters hexadecimal PIN code. The
maintainers of DVBM do not clarify why they decided to use PDFs to place the ballots.

\subsection{Tallying Process}
After the election ended, the officials transfer the encrypted ballots to a non-networked computer and decrypt it with
the offline-kept private key. This non-networked computer is used to count the votes. As a last step, the outcome is
published online on their official website.

\subsection{Security Problems}
DVBM relies on known mechanics a normal user is familiar with: Receive a postal mail with authentication credentials
informing you to vote, authenticate at a web page, download and fill out a PDF file and upload it. This makes it easy to
understand for non-technical people, since most of the people using the computer are familiar with web applications and
PDF files.\\
Therefore, it is very simple, but lacks of many security issues:

\subsubsection{Non-Encrypted Ballots}
\label{dvbm-non-enc-ballots}
The ballots are not being encrypted on the voter's machine. Only the server encrypts them to store them in the database.
These unencrypted ballots enable insider-attacks (see~\ref{insider-attacks}), because a malicious maintainer of DVBM
might easily copy, change or read the ballots.

\subsubsection{Single Point-of-Failure}
The application server encrypts the ballots. Therefore, the unencrypted ballots are accessible on this server. If an
attacker might get access to this server, she could be able to manipulate the plaintext-ballots as she wishes. So, the
ballots must be encrypted \emph{before} they are sent via Internet to prevent altering or reading the vote.

\subsubsection{Coercion}
This system does not provide any protection against coercion. Selling the credentials or being forced to vote for a
specific party is smoothly possible and can not be easily revoked or overridden as we saw it in Estonia (although even
their re-voting policy is \emph{not} a complete protection against coercion!).

\subsubsection{State-Level Attacks and Anonymity}
There are no mechanisms against monitoring / timing attacks in this system. Therefore, a state-level attacker could
observe the network traffic, which enables coercion.

\begin{table}
    \centering
    \begin{tabularx}{0.95\textwidth}{l l}
        \toprule
        \emph{DVBM}             & \\
        \midrule
        Authentication          & With postal credentials\\
        Voting Policy           & Fill out one PDF ballot and upload it unencrypted\\
        App Structure           & Web application\\
        Distributed             & Dedicated servers, but all in one data center\\
        Development Model       & Closed source\\
        Encryption Scheme       & Public key encryption, but only on the servers to\\
                                & store them\\
        Ballot Anonymity        & Unknown\\
        Tallying Process        & Separate server with offline-kept private key\\
        Voter Verifiability     & No\\
        Universal Verifiability & No\\
        \bottomrule
    \end{tabularx}
    \caption{Summary of DVBM}
    \label{table:dvbm}
\end{table}

\subsection{Summary}
DVBM tries to focus on well-known mechanics a typical voter understands. Basically, this is a good decision, but the
implementation failures make it insecure.\\
Also this system shows that encrypting the ballots on the servers and not locally is not a good design decision, because
it possibly enables more attacks (see~\ref{dvbm-non-enc-ballots}).

% Attacking the Washington, D.C. Internet Voting System
%% About
% web based
% Open Source
% secure communication over TLS

%% Voting Process
% voter authenticates with credentials provided by mail
% gets blank PDF, fills it out and sends it unencrypted back
% the server encrypts the ballots and stores them with the election’s public key
% the private key is held offline

%%%%%%%%%%%%%%%%%%%%%%%%%%%%%%%%%%%%%%%%%%%%%%%%%%%%%%%%%%%%%%%%%%%%%%%%%%%%%%%%%%%%%%%%%%%%%%%%%%%%%%%%%%%%%%%%%%%%%%%%

\section{Norwegian I-Voting System}
\label{norwayvoting}
Norway used a remote electronic voting system for the county council elections in 2011. The general design is similar to
the Estonian system, but has some other decisions regarding the design, which we are going to explain in this section.
It is presented as the first governmental voting system fulfilling the requirements for coercion-resistance and voter
verifiability. The system is developed with the help of \emph{Scytl}\footnote{\url{http://www.scytl.com/en}}, a Spanish
company which is awarded for the electronic voting systems they are developing.\\
However, in June, 2014 the Ministry of Local Government and Modernisation decided to discontinue the Internet voting
pilot project due to security concerns~\cite{NorwayEndOfficial2014}; many parts of the eligible population feared that
their votes could become public, which might undermine democratic processes~\cite{NorwayEndBBC2014}.

\subsection{Application and Voting Process}
\label{norwayapplication}
Many institutions in Norway rely on \emph{MinID}, which ``\emph{provides access to public services at a medium-high
level of security}''~\cite{MinIDWeb}. Therefore, the maintainers also rely on this service, since it provides a secure
authentication mechanism and is already accepted by the Norwegian citizen. It is free of charge and can be used to login
to the e-voting system with the voter's national identity number, a password and a PIN code.

The e-voting client is written in Java and contains the election's public key like the Estonian system and encrypts the
ballot with this public key. A signature is added with MinID and this double-envelope ballot is sent to the ballot box.

\begin{figure*}
    \includegraphics[width=\textwidth]{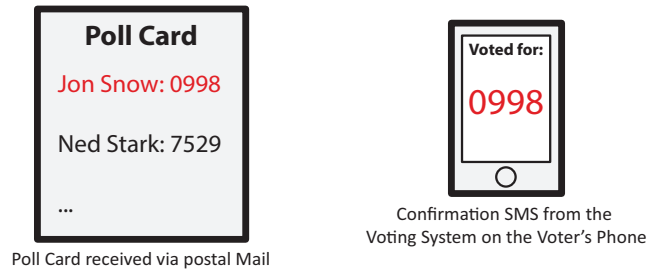}
    \caption{Compare received SMS with the Codes on the Poll Card}
    \label{fig:norwaypollcard}
\end{figure*}

Norway is different in how they implement voter verification compared to Estonia: Here, the voter gets a poll card via
postal mail containing a list of all parties and a corresponding four-digit code individually calculated for each voter.
We illustrated this in figure~\ref{fig:norwaypollcard}. After the voter placed her ballot, the voting server sends an
SMS to the voter's mobile phone containing her own 4-digit code. Now she can compare if the code she received matches
the code of her chosen party on the poll card. This gives the voter the possibility to verify her vote.\\
For coercion protection, the voter is allowed to cast multiple votes, whilst the last one counts and her previous votes
are automatically revoked.

\subsection{Tallying Process}
\label{norway-tally}
The valid ballots are sent through a mix-net (see~\ref{mixtyp}) to anonymize them with shuffling and re-encryption. This
mixing is verifiable with zero-knowledge-proofs (see~\ref{zerknopro}), like the
proof-of-correct-operation~\cite{Stenerud2012}. The proof itself is described in~\cite{Gjosteen2010, Gjosteen2013}.\\
Multiple \emph{auditors} organize the tallying process, initiate to close the ballot boxes and transfer the
ballots through the mix-net to the \emph{decryptors}. These decryptors log which anonymized ballots they received and
decrypts the ballots with the offline-kept private key and an auditor as a verifier~\cite{Gjosteen2013}.

To improve this process, Gj{\o}steen is designing a different tallying process for the Norwegian voting protocol to use
homomorphic encryption and first compress the encrypted votes and then exponentially combine the
ciphertexts~\cite{Gjosteen2013}. This might significantly reduces the computational time and makes it more suitable to
tally the ballots.

\subsection{Security Problems}
The Norwegian e-voting system has some security problems resulting in the architecture of the system:

\subsubsection{Network Attacks}
An ISP-level attacker might easily observe the traffic and create a relation between a voter and the voting servers.
Therefore, the voter is partly susceptible to coercion attacks.

\subsubsection{Centralized Infrastructure}
Centralized servers are a problem, because they are a single point of failure. An attacker needs to compromise one
server and might be able to influence the whole election.

\begin{table}
    \centering
    \begin{tabularx}{0.95\textwidth}{l l}
        \toprule
        \emph{Norway}           & \\
        \midrule
        Authentication          & MinID (ID number, password and PIN code) \\
        Voting Policy           & Multiple votes\\
        App Structure           & Client/Server with native Java app\\
        Distributed             & Dedicated servers, but all in one data center\\
        Development Model       & Closed source\\
        Encryption Scheme       & ElGamal, homomorphic encryption scheme\\
        Ballot Anonymity        & Signature stripped off from the ballot and then re-\\
                                & encrypted with a mix-net\\
        Tallying Process        & Decrypted and tallied \\
        Voter Verifiability     & Compare displayed code in application with the provided\\
                                & personalized poll card\\
        Universal Verifiability & Partly, ZKPs for shuffling published\\
        \bottomrule
    \end{tabularx}
    \caption{Summary of the Norwegian I-Voting System}
    \label{table:norway}
\end{table}

\subsection{Summary}
This voting system tries to focus on voter verifiability besides the cryptographic security. Double envelopes are used,
the mixing achieves a reliable anonymity of the ballots and the return-codes give the voter the possibility to verify
her vote. Even the mixing is verifiable thanks to the ZKPs published after the ballots passed the mix-net. These are
important steps towards a secure and verifiable voting system, but it still needs improvement in operational security,
universal verifiability and the development model.

%About
%  Java Application running on each Client
%  gets a poll card by mail containing a 4-digit code for each party, personally generated for the voter

%Voting Process
%  authentication via MinID \cite{MinIDWeb}
%  make selection inside Java App
%  Java App encrypts it
%  voter receives 4-digit return code via SMS and can compare it with the poll card
%  can cast unlimited votes

%%%%%%%%%%%%%%%%%%%%%%%%%%%%%%%%%%%%%%%%%%%%%%%%%%%%%%%%%%%%%%%%%%%%%%%%%%%%%%%%%%%%%%%%%%%%%%%%%%%%%%%%%%%%%%%%%%%%%%%%

\section{New South Wales iVote System}
\label{australia}
Australia started the world's largest deployment of an e-voting system to date~\cite{Halderman2015}. About 5\%
(approx. 280,000 voters) of the eligible citizen, placed their vote via iVote in the 2015 state election in New South
Wales.\\
This system is also developed with the help of Scytl, but the design of this voting system is much different than the
one from Norway.

\subsection{Application and Voting Process}
\label{australiaapplication}
iVote is completely closed source and provided via a JavaScript-powered website. The voter needs to register prior the
election and receives her credentials, an 8-digit iVote ID and sets her own 6-digit PIN. With these credentials, she
can login at \url{https://cvs.ivote.nsw.gov.au} and place her vote during the election. This system was designed for
eligible voters which are not in Australia during the election and unable to go to a polling station.\\
To place a test ballot, the New South Wales Electoral Commission prepared a mock election reachable under a different
URL using the same software as the real election.\\
There are three possibilities to vote with iVote: via telephone, Internet or with computers at the polling station. The
ballot is locally encrypted with a JavaScript library and then sent to the Verification Server. After this step, the
voter receives a Receipt Number to verify her vote with an automatic telephone system or online via a Receipt Server.
The voter has to enter her iVote ID, PIN and the receipt number to get a verification that her vote was properly
counted. Verification with the telephone system ends when the election stops, the online verification server stays
available after the election.

Being able to prove for which party she voted, is similar to getting a receipt and this enables coercion. This is not a
good design decision, as coercion and vote-buying are big problems in voting systems.

The description of the system itself is very inconsistent. For example, there exist several descriptions about the
encryption of the ballots: some say the system uses symmetric AES with the Receipt Number to encrypt the ballots or they
use ElGamal as the chosen scheme with the election's public key. Halderman et al.\@ found that both ways are
used~\cite{Halderman2015}. This is not the way a trustworthy voting system should be described.

\subsection{Tallying Process}
Analyzing the tallying process is not possible, because there are no publications and no source code of the application.
The Office of New South Wales Electoral Commission describes the process on the homepage as follows: ``the ballot is
decrypted, audited and counted''~\cite{iVoteOnline}.

\subsection{Security Problems}
\label{ivotesecurityproblems}
\subsubsection{Closed Source}
It is not possible to peer-review the code. This is not a good practice, because only few people with access to the
source code are able to check it for bugs or malicious code. Creating an open source voting system is essential for a
system that shall be trusted during an election.

\subsubsection{Coercion}
There is no coercion-protection in this system. A voter just needs some credentials and is able to vote. Therefore,
vote-selling and coercion are a big issue and should not be underestimated.

\subsubsection{Anonymity and Timing Attacks}
This system is again vulnerable against network-level attackers, since there are no mechanisms for anonymity provided by
the developers of iVote.

For this election, Alex Halderman and Vanessa Teague started their own independent and uninvited security analysis of
the iVote system and published their results in~\cite{Halderman2015}. Since iVote is closed source, Halderman et al.\@
could only analyze the web application, but they were not eligible to vote for this election. They analyzed the HTML and
JavaScript code used for the website and found a critical security issue.

\subsubsection{FREAK Attack}
The web application uses, as it is common, multiple servers to load the JavaScript files needed for the application. Two
weeks before the election started was the FREAK attack a big issue all over the Internet: It was possible to downgrade
the SSL/TLS encryption to 512-bit export-grade RSA on a vulnerable web server or some (older) web
browsers~\cite{FREAKAttackOnline}. This attack makes it possible to factor the RSA keys within 7 hours and about \$100
with Amazon EC2~\cite{FREAKCracking}. As a result of this downgrade, it is possible to impersonate a web server and
enables man-in-the-middle attacks.\\
In this case, iVote downloaded a JavaScript file from Piwik's enterprise service\footnote{Accessible through an
apparently secure server: \url{https://ivote.piwikpro.com}}, whose web servers were vulnerable to the FREAK attack.
Halderman et al.\@ found this issue and reported the problem to the maintainers of iVote. The Office of New South Wales
Electoral Commission reacted and dropped the JavaScript file from the insecure web server, but during this time it was
theoretically possible to manipulate the votes of the 66,000 voters, which were already placed while the attack was
possible.\\
This attack made it possible to impersonate the web servers of Piwik\footnote{Piwik is an open source tool to collect
statistics of the visitors of the website, like Google Analytics.} and serve manipulated JavaScript code, which is
for example able to alter or drop the vote. The attack attack was possible, because the impersonated Piwik web server
did not break the same-origin-policy and could therefore be used to serve a malicious JavaScript file.\\
The manipulated JavaScript file enables an attack published in~\cite{Halderman2015}, where the authors describe that it
is very easy to copy the automatic telephone system for verification, change the telephone number the voter gets
displayed on the web site and confirm a wrong ballot to the voter. Changing the website's content is very easy when the
attacker has access to a JavaScript file. Therefore, the voter might not able to notice that her vote was manipulated.

This FREAK issue has been patched during the election after Alex Halderman contacted the responsible persons.

\subsubsection{Consequences}
The system's FREAK vulnerability led to the possibility that an attacker might have exploited the vulnerability to alter
ballots, which were cast during this time. Until the FREAK issue was patched, about 66,000 votes were cast, which
potentially could have been manipulated.\\
This was a critical vulnerability, because only 3,177 votes were needed to decide to whom the last seat in the
Legislative Council was granted~\cite{Halderman2015}. Therefore, the attack might have had an impact on the results, if
an attacker exploited the FREAK vulnerability. There was no possibility to retroactively check for a FREAK attack
compromise.

\begin{table}
    \centering
    \begin{tabularx}{0.95\textwidth}{l l}
        \toprule
        \emph{New South Wales iVote System} & \\
        \midrule
        Authentication          & Multiple PINs, prior registration\\
        Voting Policy           & Unknown\\
        App Structure           & Web application\\
        Distributed             & Unknown\\
        Development Model       & Closed source\\
        Encryption Scheme       & AES, ElGamal for the ballots\\
        Ballot Anonymity        & Unknown\\
        Tallying Process        & Unknown\\
        Voter Verifiability     & Yes, via phone and the Internet\\
        Universal Verifiability & No\\
        \bottomrule
    \end{tabularx}
    \caption{Summary of iVote}
    \label{table:australia}
\end{table}

\subsection{Summary}
This voting system is purely web-based, which is an advantage, because the user does not need to install any software on
his computer. Encryption is completely realized via JavaScript in the web browser, which is no problem since these web
techniques are powerful enough for this task.\\
Lack of description of the system leads to general mistrust with the voting; there is no possibility to verify the
voting process.\\
In some cases, iVote relies on the telephone for placing the vote and for voter verification. But there is no
description about the encryption of the phone call. So, we have to assume there is none, which is a potential security
issue.\\
Summarized, the maintainers of the system seem to ignore advances and contemporary good practice used in other systems,
although the same company was involved as in the Norwegian system.

\section{Civitas}
\label{civitas}
As a last system, we would like to introduce Civitas, an open source solution for electronic voting systems based on
JCJ~\cite{Juels2005}. It is designed to be completely distributed and one of the few systems which attest themselves
coercion-freeness. This system has not been used in official parliamentary elections, but is often used as an example
for a distributed, verifiable and mostly secure voting system.\\
Civitas is developed in \emph{Jif}\footnote{\url{https://www.cs.cornell.edu/jif}}, which is a security-typed programming
language extending Java. Just as Jif, Civitas is developed and maintained by the Cornell University.

\subsection{Initial Setup}
Since Civitas is open source and well described, we can give a detailed overview~\cite{Clarkson2008}.\\
In the first step, the \emph{supervisor} creates the election, defines the parameters (re-vote policy, ballot design,
\dots), posts it on an empty bulletin board and selects her tellers by publishing their public keys. Then the
\emph{registrar} defines the eligible voters with their public keys and creates a registration and designation key for
each voter. The voter receives these keys prior to the election and the keys are based on RSA, but any other algorithm
for asymmetric key-generation could be used as well.\\
After this step, the \emph{tabulation tellers} collectively generate a key-pair, whereby voters can encrypt their votes
and credentials with the election's public key. This key-pair is based on Distributed ElGamal (as seen in~\ref{diselg})
and the public key is published on the bulletin board.\\
As a last step, the \emph{registration tellers} create the voter's public and private credentials, while the public part
of the credentials is published on the bulletin board and the private part is shared between all of the registration
tellers. All registration tellers have to cooperate to generate these credentials (see \emph{secret sharing},
section~\ref{diselg}).

\subsection{Voting Phase}
Each voter authenticates with her registration key at the registration tellers to acquire a share of her private
credential. This share can be gathered with the voter's designation key and then run a protocol with the registration
teller, which ``releases the teller's share of the voter's private credential to the voter''~\cite{Clarkson2008}. The
private credential is needed to place the ballot at a distributed ballot box and must be gathered from \emph{all}
registration tellers, whereby each registration teller has a part of the voter's private credential.\\
Now, the voter can place her vote at one of the ballot boxes. She needs to encrypt her private credentials and her
ballot. Moreover, she needs to add a well-formed proof for her ballot. Voting multiple times is possible if the
supervisor allowed it; then only the last vote counts.

\subsection{Tallying Process}
\label{civitastally}
When the election ends, the ballot boxes sign and transmit the encrypted ballots to the tabulation tellers. The tellers
verify the proofs of well-formedness and remove the faulty and duplicate ones. After this step, they anonymize the
ballots by performing random permutations through a mix-net (see~\ref{mixtyp}) and verify the private credentials in the
end in accordance to the public credentials posted on the bulletin board. Then the ballots (not the credentials) are
decrypted and counted.  Each ballot is published on the bulletin board for verification by the users.\\
Each teller publishes her proofs on the bulletin board to make each step publicly verifiable.

\subsection{Security Problems}
\subsubsection{Coercion Resistance}
\label{civitascoercion}
To create fake credentials, the voter needs her private designation key and then runs a local algorithm. These faked
private credentials are indistinguishable from the official credentials provided by Civitas, but ballots encrypted with
them will not be tallied in the last stage of the election.\\
Obvious coercion attacks are not possible with this mechanism: If an adversary wants to buy the credentials, the voter
can create fake ones and hand them out. Equally, fake credentials can be used if someone is forced to vote for a
specific candidate -- the coercer can never be sure if the real or the fake credentials were used.\\
Civitas' trust assumption 4: ``\emph{The channels on which voters cast their votes are anonymous}''~\cite{Clarkson2008}
describes Tor as a solution to provide an anonymous channel, but we demonstrated in a paper that Tor and other
low-latency networks are not enough to provide an unlinkable channel, which is needed to disable timing
attacks~\cite{TorIsNotEnough2015}. As long as this problem is not solved, even Civitas is not coercion-resistant and
coercion is still possible.

\subsubsection{Unpopular Programming Language}
\label{jif-programming}
Civitas is written in \emph{Jif}, which extends Java in the information flow control and access control. The main
problem of this dialect is in our opinion that there are only few people using this language. Even if it is used as an
open source application for an election, the number of people being able to inspect the code is too low.\\
According to the TIOBE Index of April 2015, which measures that Jif does not even appear in the top
100~\cite{TiobeApril2015}.

\begin{table}
    \centering
    \begin{tabularx}{0.95\textwidth}{l l}
        \toprule
        \emph{Civitas}          & \\
        \midrule
        Authentication          & Multiple types of credentials (fake / real)\\
        Voting Policy           & Different vote policies possible\\
        App Structure           & Java application\\
        Distributed             & Yes, except the bulletin board\\
        Development Model       & Open source\\
        Encryption Scheme       & ElGamal\\
        Ballot Anonymity        & Verifiable re-encryption mix\\
        Tallying Process        & Anonymized ballots are decrypted and published\\
        Voter Verifiability     & Yes\\
        Universal Verifiability & Yes, published ballots and ZKPs for all steps\\
        \bottomrule
    \end{tabularx}
    \caption{Summary of Civitas}
    \label{table:civitas}
\end{table}

\subsection{Summary}
The cryptographic primitives used by Civitas are well chosen and current best-practices to secure and anonymize the
ballots.  All design decisions are described in their publication and there exists an open source implementation of
it~\cite{CivitasOnline}. This makes it to one of the most interesting implementations of an electronic voting system.
But there are still unsolved security problems and the system's usability is generally very bad. It is difficult to
explain to a typical voter, what the different kinds of credentials types are for.\\
Civitas is also the first system which was from scratch designed for maximum distribution of its parts as it is based on
JCJ~\cite{Juels2005}. All parts can be multiplied and distributed on different servers and locations, except the
bulletin board. But there is no active development in Civitas; the last update was released in 2008. Therefore, Civitas
can be considered as a proof-of-concept for a distributed voting system and parts of it can be used as a base for the
development of an own system.

% last update 2008
% untappable channel
% usability
% type of application?
%   good decision to choose Jif?

% Civitas
%
% About:
% distributed trust
%
% Setup
% supervisor creates election, posts it to a bulletin board and defines the tellers by publishing their public keys
% registrar defines the eligible voters with their public keys
% each voter needs a registration and a designation key
% tabulation tellers collectively generate a public key for a distributed encryption scheme and post it on the
%   bulletin board
% to decrypt the votes, all tabulation tellers are needed to create the private key
% registration tellers create credentials for the voters.
% credentials: public and private value, while public values are posted on the bulletin board and private values are
%   shared between the registration tellers
%
% Voting
% Voters contact the registration tellers with their registration key. With a separate protocol, the designation key
%   and all registration tellers the private credential is reconstructed and given to the voter.
% then the voter can place her vote with her encrypted choice, her encrypted private credential and a proof of the well
%   formed ballot

% Coercion resistance:
% provide also fake credentials

%%%%%%%%%%%%
\newpage
\section{Comparison}
Now that we have a brief overview of the most important voting systems, we can directly compare them:

\begin{table}[H]
    \centering
    \begin{tabularx}{\textwidth}{lccccc}
        \toprule
        & \emph{Estonia} & \emph{DVBM} & \emph{Norway} & \emph{iVote} & \emph{Civitas} \\
        \midrule
        Authentication & eID & Postal & MinID & Postal & Postal \\

        \rowcolor[gray]{.9}
        Voting Policy & Multiple & Single & Multiple & Single & Multiple\\

        Application Structure & App & PDF / Web & App & Web & App\\

        \rowcolor[gray]{.9}
        Distributed & Partly & Partly & Partly & Unknown & Yes\\

        Development Model & Partly Open & Closed & Closed & Closed & Open\\

        \rowcolor[gray]{.9}
        Encryption & Public Key & Public Key & ElGamal & ElGamal & ElGamal\\

        Ballot Anonymity & Partly & Unknown & Yes & Unknown & Yes\\

        \rowcolor[gray]{.9}
        Tallying Process & \multicolumn{5}{c}{Decrypted and tallied}\\

        Voter Verifiability & Yes & No & Yes & Yes & Yes\\

        \rowcolor[gray]{0.9}
        Universal Verifiability & No & No & Partly & No & Yes\\

        \bottomrule
    \end{tabularx}
    \caption{Comparing the Systems}
    \label{table:comparison}
\end{table}

\subsection{Interpretation}
At first we notice that there are some unknown features of the voting systems. These mostly come due to the lack of a
sufficient documentation. Especially, the \emph{Ballot Anonymity} is not well described in two of five systems, but this
is one of the most critical issues in a voting system. If a voter can not trust that her vote is anonymous, she could
not trust in the system.\\
It is conspicuous that each system tries to fulfill the preliminaries on their own way. Mostly, they only agree in the
tallying process. Voting systems used in big elections, are only partly distributed. This means that there are often
different machines for different tasks (firewall, authentication server, mix-net servers, etc.), but they are all
deployed in one data center, which makes them vulnerable against DDoS attacks and this risks the availability.

\paragraph{Authentication}
Estonia provides the easiest authentication method, since it uses their eID cards, whose technique is well known and
understood. Norway tries to go a similar way with MinID. The other systems use the postal mail, because this is assumed
to be the easiest and most trustworthy way to provide the eligible users their credentials.

\paragraph{Voting Policy}
Most systems rely on multiple votes. This is natural, since this is the easiest way to make coercion much more difficult
and less effective. But no system provides a complete secure solution against coercion.

\paragraph{Application Structure}
Since web-techniques are powerful enough, native and web applications are equally used. As long as the developers
deploy the application for the common operating systems and mobile devices, the eligible voters are able to access the
application and can therefore participate in the election. But native applications are not as easy to use as web
applications, because they need to be distributed to the eligible citizens and then installed.

\paragraph{Encryption Scheme}
Three out of five analyzed systems use ElGamal as their encryption scheme, but none of them use the homomorphic
addition of the ciphertexts, because the homomorphic aggregation of the ballot takes a long time and much computing
power (see subsection~\ref{reencryption}). ElGamal is often used because of the possibility to use threshold encryption or
the re-encryption property.\\
The ballots of these systems are decrypted and then tallied the ``normal'' way. One reason is the
computational power needed to directly tally the ballots on their ciphertexts~\cite{Clarkson2008}. Therefore, they are
first decrypted and then counted and not directly aggregated on their ciphertexts.

\paragraph{Ballot Anonymity}
This is a critical part: In three out of our five systems, it is either unknown or not sufficiently guaranteed that the
ballots are anonymized. These flaws are a relevant criterion for exclusion of these systems, since they do not fill the
most important preliminary for a voting system: to keep the voter's choice reliably anonymous.

\paragraph{Voter and Universal Verifiability}
Most systems provide the possibility to verify the voter's ballot, but the system's tallying processes are not
completely verifiable. Only the Norwegian system provides zero-knowledge-proofs to verify the mix-net.

\subsection{Summary}
None of the analyzed real-world system fulfills all preliminaries defined in chapter~\ref{preliminaries}. Nevertheless,
these preliminaries are essential for a secure, anonymous and verifiable voting system. This is not satisfying for
critical elections and we can already say at this point that currently implemented voting systems are not sufficient
for parliamentary elections.

It seems that some implemented voting systems ignored the research and experiences, which other systems prior to them
had. We can just pick South Wales iVote, which started the most recent state election in 2015. It looks like they tried
to reinvent electronic voting and implement many things differently compared to older voting systems. Some major flaws
are that they completely ignore coercion, enable vote-selling, do not document the election's process and provide no
universal verifiability.\\
The Norwegian system and iVote are developed by the same Spanish electronic voting company \emph{Scytl}. Since the
Norwegian system was developed \emph{before} iVote, this company should have the experiences to know better how to
reduce coercion or even how to implement a mix-net for reliable anonymization of the ballots. We discuss this later in
section~\ref{scytl-company}.

%%%%%%%%%%%%%%%%%%%%%%%%%%%%%%%%%%%%%%%%%%%%%%%%%%%%%%%%%%%%%%%%%%%%%%%%%%%%%%%%%%%%%%%%%%%%%%%%%%%%%%%%%%%%%%%%%%%%%%%%

\section{Other Systems and Schemes}
The system of this chapter is only a small collection of existing e-voting systems. There are several theoretical
schemes or systems for low-coercion elections, but most of them lack of good usability or are too complicated and are
therefore just used as a proof-of-concept.

\subsection{Helios}
\label{helios}
Helios is a web-based voting system developed for small elections, like a student's parliament in a
university~\cite{Adida2008}. It provides no mechanisms against coercion and could therefore not be used in parliamentary
elections, which is the reason why it appears in this section. The complete application is open source and the
credentials are sent via email to the voters. Good practices are taken into concern and are correctly implemented,
e.g.\@ ElGamal is used and the ballots are shuffled and re-encrypted in a mix net.\\
All steps are verifiable; the cast votes, ZKPs of the mix-net, shuffled votes, decrypted votes and the
proofs-of-correct-decryption are published on a bulletin board.

\paragraph{Test-Audits against Ballot Modification}
A special feature of Helios is the possibility for auditing the own vote. After composing the ballot, Alice can choose
to send her vote to the Helios servers or reveal the used random factors needed for encryption and show the ballot's
content. This step is also possible on her local machine with a provided Python script, whereas no Internet connection
is needed and the ballot is destroyed after the auditing process.\\
Being able to audit her vote makes it possible to verify that no malicious software has modified her ballot. She can
repeat these steps as often as she likes until she decides to send the ballot to Helios. This mechanism makes it
unlikely for a malicious program to modify the ballot without noticing, since this program does not know if Alice casts
her vote or keeps it for auditing.\\
A voter has also the option to check if her vote was modified. Her encrypted ballot is locally created and posted on the
bulletin board combined with her name. Therefore, she could compare the encrypted ballot from her device with the posted
ballot on the bulletin board and might reveal the modification of her vote.

These test-audits might be a good proof-of-concept to reveal modifications on the own ballot, but are not usable in
practice, which was the result of a usability analysis~\cite{Karayumak2011}.

\subsection{Code Voting}
Several e-voting systems follow a different approach for verification. These systems belong are using Code Voting, which
is an enhancement to the poll cards we already saw in the Norwegian I-voting system in subsection~\ref{norwayapplication}.
To vote, a ballot and a unique code card are needed, whilst the coding card is shipped to the voter through an
untappable channel. The candidates on the ballot are shuffled and marked with a random number and the coding card has a
unique serial number. Extensions to this are more complex and are coded into matrices. These coding cards are some kind
of $n \times n$ matrices with a row $n$ for each candidate, where, for example, each row contains one cell with ``yes''
and all $n-1$ remaining fields might be marked with ``no''~\cite{Kutyowski2010a}.\\
If the ballot and the coding card are put side by side, she can mark the ``yes'' cell in the row of the candidate she
wants to vote for and one random ``no'' cell in the rows of the other candidates. Giving the system her marks and the
coding card's serial number, the system is able to tally her vote.\\
Each voter gets one ballot, but might have many coding cards. After she placed the ballot, she can get a receipt,
wherewith she can verify her vote afterwards. The receipt is useless without the ballot and the correct coding
card.

Popular systems using code voting achieve to provide end-to-end verifiability with this new kind of receipts and coding
cards. Some examples are \emph{Punchscan}~\cite{Punchscan2009}, \emph{ThreeBallot}~\cite{Rivest2004} or more advanced
\emph{Scratch, Click \& Vote}~\cite{Kutyowski2010a}.

\chapter{Construction}
\label{construction}
Constructing an electronic voting system is a very complex task. High requirements for integrity and anonymity do
not allow a simple solution; each building block must be well chosen. Real-world internet voting systems have shown
that it is very easy to do something wrong: Possible vote manipulation (see~\ref{estonia}, \ref{australia}) or dubious
storing of the votes (see~\ref{dvbm}) lead to low trust in electronic voting systems. This is comprehensible, since
poorly designed voting systems endanger the anonymity of the voter's choice or even threaten democracy through the
possibility for an attacker to have an impact on the election's outcome.

With the analyzed systems in mind, we try to design a voting system which fulfills the requirements. In the end we
will show which issues can not be solved and need further work to find a solution. We already described the assumptions
for our system in section~\ref{assumptions} and add another assumption in subsection~\ref{eid-assumption}.

\section{Registration and Authentication}
\label{registration-authentication}
As a first step, the voter needs to register for the election to get her credentials needed for authentication. As we
have seen in chapter~\ref{systems}, there basically exist two options: credentials via postal mail or an electronic ID
card with a governmental controlled PKI.

Since more and more countries are introducing electronic identity cards (eID), it is just natural to think about using
it for authentication in electronic voting systems. Governments with electronic ID cards control the electoral register
and provide the public-key-infrastructure needed for authentication with asymmetric cryptography
(see~\ref{estoniapki}).\\
This is a great starting point for our voting system, because the voter can authenticate herself at the authentication
servers and it is easily possible to look up the electoral register to see if the voter is eligible to vote.\\
Furthermore, vote-buying becomes less relevant, since a voter would reconsider if she sells her electronic identity,
which might be used for several online services. Selling some one-time-credentials for one specific election is much
easier than giving up her own identity manifested in the eID card. This is an example for \emph{expensive credentials},
which can be used for other services and are therefore more unlikely to be sold.

An alternative to the eID card are credentials sent via postal mail to the eligible voters. These credentials can be of
two different kinds: Firstly, they contain the codes to log into the voting application and place the vote. But this
would easily enable vote-buying. Secondly, credentials could be sent to the voter giving her the option to register her
own public key, which also enables vote-buying as described in the previous paragraph. Basically, this could be as
secure as the electronic ID card and has the advantage that the voter can create the key-pair herself. However, both
methods require a second trustworthy channel \emph{and} a trusty government-hosted PKI. Therefore, sending postal mails
do not provide any relevant advantages compared to eID and achieve the same level of trust, which is why we chose a
secure electronic ID card in our trust assumptions.

\subsection{Assumption}
\label{eid-assumption}
Electronic ID cards are heavily discussed in the community and criticized for the general practice and many possible
security problems, like compromised systems with which an attacker could access an eID when it is plugged into the card
reader~\cite{CCC2013}. But this is not in the scope of this thesis. Therefore, we assume that we have
\textbf{trustworthy ID cards} and are able to \textbf{securely authenticate} at our e-voting application with a secure
(government-managed) \textbf{PKI}.

This also makes the registration dispensable, because of the governmental managed PKI, the voting system can verify the
eligibility with the data from the eID card or with an API request to a server provided by the government containing the
electoral register.

\subsection{Using eID cards}
From the real world systems, the Estonian one seems to be most advanced. They are actually using their eID cards to
authenticate at the voting system, as we described in section~\ref{estoniapki}. This is one of the easiest solutions for
authentication, since the citizens are already familiar with this technique.\\
Most other systems assume that there exists at least one untappable channel (a secure channel, where messages can be
sent without attackers having the possibility to eavesdrop or manipulate the content), which they use to send
credentials to each eligible citizen. In times of mass-surveillance it is questionable if there really exist some
untappable channels like the postal mail (which is often assumed), but with eID cards we can skip this channel and use a
secured connection in the Internet to connect and authenticate at the voting application.

If there are no PKI and eIDs provided by the government, the voters should be able to get their credentials from an
authority which they can personally visit, identify themselves and get the credentials for the election. Sending them
via postal mail might be another option, but we are sceptical that this is really a trustworthy channel, since the
Snowden leaks proved that the NSA tracks the snail mail in the USA~\cite{Forbes2013}.

%%%%%%%%%%%%%%%%%%%%%%%%%%%%%%%%%%%%%%

\section{Coercion Freeness}
\label{coercion-freeness-constr}
Depending on the building block chosen for the authentication, the susceptibility for coercion can be estimated. For
example, it is easier for someone to sell her credentials, which are only needed for this election, than to give the
complete own electronic identity to a coercer, because with the eID the coercer might be able to sign documents in the
voter's name but for his behalf or similar. What we try to achieve in the voting systems is coercion-freeness.

As already described in the preliminaries, an e-voting system must be coercion-free. This means for a voting system that
a voter must not be able to prove to the coercer how she voted. In our opinion, the definition of coercion should
include that even network attackers are not able to show \emph{that} the voter placed her ballot or not. This stricter
definition of coercion seems natural, since we know that network- or ISP-level attacks are possible. Resultant problems
are that this observer can force groups of voters simply \emph{not} to vote, e.g.\@ because this group of people is
known for always voting for party A. This kind of attack is currently not considered by most voting systems, because
there were no such attacks documented as far as we know.

Some modern electronic voting systems allow multiple votes in their policies to reduce coercion
(see~\ref{estoniavoting},~\ref{norwayapplication}). This way, the voter is able to place a new ballot which revokes the
old one, which is an easy to use and easy to understand approach against coercion. The developers of Civitas chose a
different way and give the voter the option to create fake credentials, wherewith the voter can vote in the presence of
a coercer (see~\ref{civitascoercion}).

A complete, and still usable, protection against coercion is currently not available. Current implementations provide
only basic protection against vote-buying and parts of coercion.

\subsection{Coercion Evidence}
\label{coercion-evidence}
One of these theoretical approaches is the idea of coercion-evidence: if a certain amount of ballots are being revoked,
the election is supposed to be compromised and should therefore be closed~\cite{Grewal2013}. The authors of this
evidence assume that not too many voters use double voting to revoke their first choice as long as they are not being
coerced. But if a defined threshold is reached, there must be a coercer compromising the election and has a big enough
impact on the outcome. This threshold should be chosen as the minimum amount of votes needed to change the distribution
of the seats in a parliament, for example.

\subsection{Reducing Coercion}
\label{construction-reduce-coercion}
The theoretical approach of coercion-evidence can easily be integrated in existing voting systems providing the
possibility to replace their votes. Depending on factors we described in~\ref{coercion-evidence}, the election officials
should define an appropriate threshold.\\
Combined with existing practices as they are used in Estonia, Norway or Civitas, the officials get an easy instrument to
estimate the trustworthiness of the election's outcome.

We suggest to allow \textbf{multiple votes} for an election, whilst the last ballot counts and the paper ballot also
overrides the electronic ballots. Combined with a meaningful \textbf{threshold}, the system gets an easy-to-use
estimation mechanism to evaluate, when an election is possibly coerced. Also the suggested use of \textbf{expensive
credentials}, like the eID, make it unlikely that the eligible voter might sell her vote as described
in~\ref{registration-authentication}.\\
Further research is needed in this point to develop a more secure approach against coercion. We discuss this later
in~\ref{coercion-open-issue}.

%%%%%%%%%%%%%%%%%%%%%%%%%%%%%%%%%%%%%%

\section{Application Architecture}
As we have defined how to register and authenticate, we can now describe how the application could look like and which
blocks from existing systems can be taken.\\
Real-world voting systems are realized in two ways: as a web application, where the voter directly connects with her own
browser to the application, or with native software products the voter needs to download before the election begins.

\subsection{Web Application}
Deploying it as a normal website has obviously the advantage that a user does not need to download and install
software. Web technologies, like JavaScript, are powerful enough to locally encrypt the ballots in the browser and then
transfer it to the server. Therefore, it is possible to create a complete web-based application.

\subsection{Native Applications}
A native application can easily be verified via a checksum after downloading it, which ensures the integrity (provided
that it is possible to get the official, unchanged checksum from a trusted source). Then the election officials can ship
the election's keys or similar needed credentials directly with the application.\\
The checksum itself is \emph{not} a complete protection against infected software, but as long as a secure cryptographic
hash function (e.g.\@ SHA2) is used, we can assume that integrity is granted.

\subsection{Web vs. Native Applications}
Since man-in-the-middle attacks are likely in the Internet~\cite{SuperfishGuardian} and web servers have many weaknesses
\cite{FREAKAttackOnline}, it appears to be more secure to provide a native application, but these applications are also
provided through the Internet. Given that the users verify their software with the checksum, we can be sure that the
provided keys are the correct ones needed for the election. That is the Estonian way to prepare an election. They
develop software for all big platforms and describe very well on their websites how this software can be used.\\
The biggest problem in this part is that the typical user does not verify the downloaded application with the checksum
or even knows how to do this.

One problem is that this native software is mostly provided via the Internet and common web servers. So, there exist
the same problems in distributing the software via the Internet compared to the direct access at a web application.

Both approaches have advantages and drawbacks dependent on their implementations. Since the distribution of a web
application is more easy and no installation is needed on the voter's device, we suggest to design a \textbf{web
application}.

%%%%%%%%%%%%%%%%%%%%%%%%%%%%%%%%%%%%%%

\section{Distributed Infrastructure}
\label{distributed-infrastructure}
Each presented system, which was used for big political elections, is centralized in one data center (as far as we
know). There are mostly several servers for dedicated tasks, e.g.\@ for the firewall or the voting application. But as
some attacks have shown, when one of these servers gets compromised by an attacker, usually she has control over the
complete election or sufficient access to the ballots~\cite{Wolchok2010, Halderman2015}. In this scenario, taking over
one server is enough to have a great impact on the results of an election. Even applying a DDoS attack will use the full
capacity of the server, which is the target of this attack and it might heavily reduce the availability of the voting
servers. But availability is a preliminary for an election (see chapter~\ref{preliminaries}).

The only nearly complete distributed system we described in chapter~\ref{systems}, is Civitas
(see~\ref{civitas}). Each task needed for the election is separated in the source code and can be deployed on
different machines, no matter where they are located.

\subsubsection{Advantages of Distributed Systems}
\label{advantagesdistsystems}
\begin{quote}
    \emph{``A distributed system is a collection of independent computers that appear to the users of the system as a
        single computer'' (Andrew Tanenbaum, 1994)}
\end{quote}

This quote describes very well what we try to achieve for the architecture of a voting system: Invisible in the front
end, distributed in the back end. Distribution leads to a better availability and reliability, because of the simple
fact that there are more servers serving the application. It can also be faster than a single server structure, since
the requests can be load-balanced and it scales more easily.

Availability is a very important preliminary for a voting system, because a system is worthless if it can not be
accessed. Therefore, we must protect electronic voting systems against a lack of availability, which is nowadays mostly
caused by DDoS attacks. Distributed servers are more robust against this common attack and provide a higher reliability
than centralized solutions.\\
Naturally, the developers should also consider to use crypto-puzzles, since this is also a good protection against DDoS
attacks, but distributing the servers is also necessary to keep the voting system reliably online.

Distributed servers have also the advantage that the security increases, because the voter does not need to rely on
exactly one server, which might have been compromised. Instead it might be possible to choose the ballot box, which
provides a sense of security to her.

Details about the infrastructure of the described systems are often not publicly available. But while designing a
voting system, it should be considered to \textbf{distribute and replicate it over multiple machines in different data
centers} to gain the advantages shortly described in~\ref{advantagesdistsystems}.

Distributing the servers into several locations is one possibility to achieve these goals. But since the invention of
Bitcoin, there is also another approach, which is heavily discussed in the community: a blockchain.

%%%%%%%%%%%%%%%%%%%%%%%%%%%%%%%%%%%%%%

\section{Different Approach: A Blockchain}
\label{blockchain-voting}
With the help of a distributed database, called blockchain, it might be possible to build a peer-to-peer voting system
which does not need servers controlled by the government and distributes most of the trust and computational power
needed for the election to volunteers. These volunteers can be anyone, who want to support the election. The more people
are supporting, the more secure becomes a blockchain.\\
This database was first presented in 2008 by Satoshi Nakamato as the basic element the Bitcoin protocol was built upon
\cite{Nakamoto2008}. Currently, the Bitcoin protocol is used for the biggest peer-to-peer based payment system and was
developed as an alternative to conventional money, which is centralized and regulated, whilst Bitcoin is not.

\subsection{Definition: Blockchain}
A blockchain is a distributed database, where the complete data is shared between all participants in the network.
Data, which is supposed to be stored in this database, is packed into packages with a defined maximum size and verified
with a specific hash. This hash must begin with a certain number of zeros, which depends on the number of participants
in the network. To achieve this, the participants add a nonce to the packed data and try to find the correct hash by
modifying the nonce. This proof-of-work is called \emph{mining}. Mining is used to generate new Bitcoins. A discoverer,
who found the correct hash for some packets, is granted some Bitcoins. The amount of these granted coins is
controlled by the Bitcoin protocol and it takes about ten minutes to discover a new hash.\\
Data in the Bitcoin protocol is represented as \emph{transactions} between two or more users. Since all transactions are
public, each user knows the current amount of Bitcoins of all users. Before the transactions are added to the
blockchain, the inputs of the transactions are checked and it is ensured that these inputs have not been spent before.
This verification is possible due to the public transactions stored in a blockchain and it prevents
\emph{double spending} of the coins.

Volunteers can participate in the network in two ways:

\begin{enumerate}
    \item Set up a full node, which means to have the complete blockchain locally stored. These nodes verify the hashes
        of the blocks, which ensures the blockchain's integrity. They also exchange the blockchain among themselves to
        keep a common state. It is necessary to have an active Internet connection to participate.
    \item Miners collect new transactions and find new hashes to generate new blocks. If a hash is discovered, it is
        spread in the network and can be verified by the full nodes.
\end{enumerate}

This is the general idea behind a blockchain and the first protocol using it. As it is open source, developers can use
different approaches to verify the blocks which should be stored in the database and you can also choose a different
proof for this verification. A better suiting proof for voting systems is presented in
subsection~\ref{proof-of-stake}.

In the Bitcoin protocol, a blockchain is used to store the transactions between the users. Each user has a private
wallet and is able to buy Bitcoins, which she can then spend. All transactions are publicly available, starting with the
genesis block, which is the first block ever mined. The next block is appended to the genesis block, the third to the
second one and so on. This creates a long chain of blocks, which is then called blockchain.\\
It is possible to append the new found block to the blockchain wherever the user likes, but the protocol's design
defines that the longest chain is the ``true'' chain. Smaller chains are ignored. Combined with a timestamp and the
proof-of-work, this prevents double-spending.

\subsection{Advantages of a Blockchain}
A blockchain has several advantages, which make it a robust and secure alternative to other databases:

\begin{itemize}
    \item Completely distributed with many nodes storing the complete database.\\
        $\Rightarrow$ \emph{high availability}
    \item Each block is verified and appended to the blockchain. Therefore, it is hard to alter an older value in the
        chain, since all following blocks have to be recalculated, which needs much computational power due to the
        proof-of-work. This is also why double-spending is unlikely.\\
        $\Rightarrow$ \emph{verifiability and integrity}
    \item Easy to define one common starting point, where to store the data -- always append it to the last block in the
        longest chain.
\end{itemize}

These advantages make a blockchain attractive to be used as the database in an electronic voting system, because it
already provides a mechanism to fulfill some of our preliminaries.

\paragraph{Compare a Blockchain with a typical peer-to-peer Application}
Consider a peer-to-peer voting system, where all transactions are distributed among all online nodes. The nodes have
equal roles, since we are thinking about a voting system without the need of governmental controlled super nodes. How
are new votes handled? Typically, they are somehow broadcasted into the network, depending on the underlying routing
policy. But since the nodes do not know all participants in the network, it is not possible for them to directly
broadcast the ballot. Therefore, other nodes are needed to handle the new transaction and to forward it to their
neighbors, hoping that the ballot arrives at all nodes. This is not likely, since malicious nodes or nodes, which go
offline, might not forward the ballot. Eclipse attacks are also possible, which prevent the data exchange of one
specific node by placing malicious nodes in strategical good positions, which can then compromise the routing and
distributing the data.\\
This leads to the problem that all nodes have a different number of ballots locally stored and a common starting point
can not easily be found.

Ballots which are broadcasted into the network, are also not verified and can be spread across the complete network
without big effort. This leads to an uncontrolled flood of ballots, whereby each active node keeps a different
state.

These are some problems that can be solved with a blockchain, since many ballots are packed into one block, which are
first verified and then appended to the blockchain. The proof-of-work for each block ensures integrity of the ballots,
because of the crypto-puzzle, which needs to be solved. Also, the definition that the longest chain is the current
state of the database, makes it easy to solve this issue.\\
So, a blockchain solves two major problems when designing a peer-to-peer network with a distributed database:
maintaining one common state of the database and the integrity of the database.

\subsection{Voting in a Blockchain}
\label{voting-in-the-blockchain}
This approach is pretty interesting for electronic voting systems, because it is secure, reliable, completely
verifiable and distributed. And the computational power needed to verify the ballots, is distributed among all
participants, who want to contribute to the system.\\
There are already some organizations trying to build a peer-to-peer based voting system with a blockchain, like
VoteCoin~\cite{VoteCoin2015} or BitCongress \cite{BitCongress2015}, but their websites are partially under construction
and they have nothing published explaining their concept in detail; they only advertise the blockchain-approach, but do
not explicitly present their ideas. Therefore, we can not analyze how they use a blockchain for voting. This might be
the case because of many problems a blockchain has when it comes to voting, which we want to discuss briefly, when we
try to describe a simple blockchain voting protocol.

\subsubsection{Design}
Two approaches are thinkable: designing our own blockchain protocol or using one big and stable protocol, like Bitcoin.
For now, we define our own protocol and call it \emph{BallotCoin}, which can be used to cast one vote for a candidate.
This is more appropriate, since we want to use a different proof to verify the ballots. If we would decide to implement
it directly into the Bitcoin network, it would produce more problems like non-fitting proof-of-works or the
unpredictable progression of the Bitcoin price, since it would heavily increase, when a government buys many Bitcoins.\\
Let us transfer the elements of a voting system to a blockchain-based voting system.

\paragraph{Authentication and Registration}
Each candidate and eligible voter needs her own private wallet. The voters should be able to create their own wallets
and register them at some place to verify their eligibility. After the registration, the voter must receive a BallotCoin
to place her vote.

\paragraph{Setting up the Election}
The election's officials have to define the candidates and set up the election. One address of each candidate must be
publicly available, where the voter should transfer her BallotCoin to. Ideally, this is nicely packed into one
application, so a user only has to click on the candidate she wants to vote for.

\paragraph{Voting Process}
Voting for a candidate is related to a transaction in the Bitcoin protocol: the voter sets up a transaction transferring
her BallotCoin to the wallet of the candidate. The amount of BallotCoins in the wallet of each candidate is the result
of the election.\\
Transferring the BallotCoin to an address not belonging to a candidate should be prohibited. By this way, only real
votes are stored into the blockchain.

\paragraph{Verifiability}
The general idea of a blockchain defines that all transactions are public. This means that each user can see if her vote
has arrived in the candidate's wallet. Also, all other transactions can be verified this way to reconstruct the results
of the election.

\subsubsection{Problems}
The idea of distributing the complete voting process and making it verifiable for everyone is very interesting. But
there are several problems in this design of a voting protocol.

\paragraph{Coercion and Receipt Freeness}
All transactions are public in a blockchain. This means that a voter has a receipt of her vote: she knows her own
address and can prove to a coercer to which address she transferred her BallotCoin. Therefore, this receipt enables
coercion.\\
Moreover is re-voting not possible in this approach, because each voter only has exactly one BallotCoin. A possible
solution would be to transfer multiple coins to each eligible voter and subtract double votes when the election ended.

\paragraph{Fairness}
Partial results should not be allowed in a voting system (see chapter~\ref{preliminaries}). This could influence voters
who have not voted yet, since they are more likely to vote for one of the parties which have currently more votes. These
voters might recognize that the small party they might vote for has no chance to get a seat in the parliament; her
vote would then be wasted.\\
As long as a blockchain is publicly available during the election, fairness is broken and this should not be the case
in a parliamentary election. This broken preliminary is sufficient not to use the blockchain approach for voting.

\paragraph{Proof-of-Work not suitable}
A typical proof-of-work, like in the Bitcoin protocol, is not suitable for a blockchain voting protocol, because mining
needs much computational power and thus monetary resources to find the correct hash. It is also thinkable that
organizations with complete data centers or mining pools use their superior computational power to find the correct
hashes faster than other volunteers. This is a problem, because if a mining pool has the power to provide 51\% of the
network's computational power, it is very likely that this pool discovers most of the hashes. So, they can define the
longest chain, since they decide where to append the block to construct the longest chain. This is called
51-percent-attack, which enables double spending of the BallotCoins and endangers the integrity of the blockchain,
because one pool has a too high impact on the generation of the blocks~\cite{Hobson2013}. In 2014, the Bitcoin
mining pool \emph{GHash.io} reached 51\% of the networks hash-rate and was theoretically capable of unobtrusively
manipulating the blockchain of Bitcoin~\cite{Coindesk2014}. They announced to reduce their hash-rate to preserve the
blockchain's integrity.\\
We have to think about another proof, which is needed to secure the ballots in a blockchain.

\paragraph{Modified Proof-of-Stake}
\label{proof-of-stake}
One approach could be the \emph{proof-of-stake}, which is already used in several Bitcoin-alternatives like
Peercoin~\cite{King2012} or Nextcoin (Nxt)~\cite{Nxt2015}. The main difference is that the power to verify new blocks
in a blockchain depends on the amount of coins you have in your wallet. The total number of coins is fixed in these
systems and you earn money with transaction fees.

We focus on the Nextcoin protocol, since it provides a quite usable proof-of-stake, which we modify in the next
paragraph to make it suitable for a voting system.\\
Let us assume there exist 100 Nxt coins in our network and we own one Nxt coin. Each minute, a random node is picked by
the Nxt protocol and selected as the next miner for the new block. If this node is connected to the Nxt network, it can
forge a new block. If not, another node is randomly chosen. Forging a block means to collect open transactions, make a
new block, receive the transaction fees and spread the new block to the complete network. To increase the probability of
being picked as the new miner, it is sufficient to increase the number of coins we have in our wallet. We assumed to own
1 of 100 Nxt coins, so the probability of being picked is:
\begin{align}
    \frac{\mbox{Nxt coins we own}}{\mbox{total number of Nxt coins in the network}} = \frac{1}{100} = 1\%
\end{align}
Therefore, we have a 1\% chance of being picked by the network to forge the new block.

\paragraph{Application of the Modified Proof-of-Stake in BallotCoin}
In our BallotCoin protocol, the number of coins each eligible voter possesses should not matter. So, the chances of
being picked is equally distributed between the eligible voters. To guarantee the integrity of our blockchain, we need
some full nodes staying online during the election. These nodes verify existing blocks in the chain and might get picked
by the network to forge the new block of ballots. The picked node has then the task to verify the new ballots before
they are appended to the blockchain. If many nodes are online, it is unlikely for a malicious node to get picked, which
also secures the network.\\
Since we left computational power behind us to secure our blockchain and to keep its integrity, it is also possible for
small devices to be a BallotCoin node. This means that even smartphones and tablets can be used to forge new blocks,
given that they have the BallotCoin application installed, an active Internet connection and enough space to store the
blockchain. Scanning the network for new ballots and creating a new block is an easy task, which can be handled with
their small processors. Therefore, a volunteer in the election could just keep the application running to support the
election without spending more money than she is used to.

\paragraph{Calculating the disk usage of the BallotCoin blockchain}
To estimate the disk usage of the blockchain in a e-voting system, we focus on the Estonian parliamentary elections in
2015, because this country already provides reliable statistics about the eligible voters and the percentage of voters
already using their I-voting system. We focus on these statistics from their most recent election for the calculation
below~\cite{EstoniaStatistics2015}.\\
A typical small transaction (1 sender, 1 receiver, no transaction fees) in the Bitcoin network is about 200 bytes
long~\cite{BitcoinWiki2015}. In the last Estonian election were 176,329 I-votes counted, which lead to this estimated
disk usage for the transactions:
\begin{align}
    200 \mbox{ bytes} \cdot 176,329 = 35,265,800 \mbox{ bytes} \approx 33.6 \mbox{ MiB}
\end{align}
We have to add some overhead for the hashes for each package of transactions, but this will not heavily increase the
calculated disk usage. Therefore, if the blockchain would have been used in Estonia in the last election, the disk usage
for all transactions could have been about 33.6 MiB. Since smartphones have several gigabytes of internal storage, this
should not be a problem for most devices.

Estonia is a very small country compared to e.g.\@ Germany with 64,4 million eligible voters~\cite{StatBundesamt2014}.
Therefore, such an election would yield in Germany a much higher disk usage. But we have no statistics of the
amount of voters using the electronic voting system, which is why we rely on Estonia in this example.

\paragraph{Attackers and the Proof-of-Stake}
As all nodes are assumed to be equally qualified to forge new blocks, an attacker needs to compromise many eligible
voters or buy their identities to set up many malicious nodes. But as long as a high number of volunteers actively
participate in the network (i.e.\@ by being a full node), an attacker needs to have a huge amount of these malicious
nodes to have an impact on the integrity of the blockchain. This is not impossible, but it is very costly, since an
attacker needs many nodes to have a high probability of being picked.

It is uncertain whether enough citizens participate in the network to secure the network, because we do not believe that
enough voters stay online during an election to reduce the probability for an attacker to get picked for forging. As
long as we have not implemented this peer-to-peer system, we can not say for sure how this electronic voting system is
accepted by the votership.

\subsection{Summary}
\label{blockchain-summary}
A blockchain might be an approach to secure electronic voting. But there are nearly the same problems the centralized
voting systems also have, whilst the biggest issues are receipt-freeness, the resulting vulnerability against coercion
and fairness.

We found no e-voting systems, which follow a peer-to-peer approach except the two companies described
in~\ref{voting-in-the-blockchain}, but they do not provide any source code or concepts of possible designs.

For now, there are too many issues to provide a fully functional blockchain voting system, which fulfills all
preliminaries. But it is still a great idea to distribute the complete voting- and tallying-process over the whole
range of eligible voters.\\
Therefore, we should keep an eye on this development, but for our constructed system we suggest to use different
locations and redundant servers to distribute the voting system as described in
section~\ref{distributed-infrastructure}.

%%%%%%%%%%%%%%%%%%%%%%%%%%%%%%%%%%%%%%%%%%%%%%%%%%%%%%%%%%%%%%%%%%%%%%%%%%%%%%%%%%%%%%%%%%%%%%%%%%%%%%%%%%%%%%%%%%%%%%%%

\section{Logging Events}
\label{logging}
All events should be logged on a log server. These events include for example:

\begin{itemize}
    \item New ballot: when a new ballot is placed, create an anonymized entry
    \item Transferring data: when data is transferred between servers
    \item Log in: when someone tries to log in
    \item Other interesting interactions
\end{itemize}

This is a mechanism to monitor the data and traffic between the servers or to a server. Attacks might be revealed with
this server even when the attacker is inside the data center. At least this creates a trail if the election has been
compromised, e.g.\@ when an unusual behavior is observed. Then the election's authorities can cancel the election or
take further steps.

Restricted access to the logging server is essential. It must be ensured that an attacker is not able to compromise
the logging server. Otherwise manipulation of the logs is likely, which hides possible manipulation of the ballots.

The log server should not contain sensible entries. Otherwise the entries can not be published, because the log entries
would reveal too much of the private information of the voters.

Estonia implements this for their Vote Forwarding Server when the ballots are moved to the Vote Storage Server. Both of
these servers keep a log on the Log Server documenting what happens with the ballots during the election. Since this is
a low expense mechanism, a dedicated \textbf{log server} should be implemented, because it increases the possibility to
reveal attacks and keeps track of the ballot's movements inside the voting system itself.

%%%%%%%%%%%%%%%%%%%%%%%%%%%%%%%%%%%%%%

\section{Development Model}
Following an open source philosophy might have a big impact on the trustworthiness or security of a voting
system, which is described below. Both development models, namely open source and closed source real-world voting
systems, are used and provide different problems and advantages.

\subsection{Closed Source}
Some systems, like the South Wales iVote (see~\ref{australia}) or parts of the Estonian system (see~\ref{estonia}), do
not publish their source code. The maintainers of the Estonian system justify themselves that it would be too easy to
copy the complete system and set up a malicious one. Halderman et al.\@ took the open source parts of this system and
reverse engineered binary scripts, which were closed source, to build a working copy for testing
purposes~\cite{Halderman2014}. This copy nearly looked like the original system and had the same functionality, although
some parts of the source code were not published.\\
Since it is not possible to review the code, obvious security issues cannot easily be exploited. It is indeed a bit more
complicated to reconstruct the software and build a copy of it, but this might not stop an attacker since it is still
not too complicated to recreate the interface of an application.

\subsection{Open Source}
Many other systems, like D.C.\@ Digital-Vote-by-Mail (see~\ref{dvbm}) or Civitas, decided to make their code publicly
available. This is a big step towards transparency, since everybody is able to comprehend each step of the election or
even set up her own mock election.\\
Popular open source applications have the advantage that the community is willed in reviewing and analyzing the code.
In the case of electronic voting systems, many research groups, like Halderman et al., are also very interested in
investigating the system's structure and security. With the support of volunteers, many security issues can be reported
and solved. This leads to a more secure software design.\\
In the first prototype of DVBM, the research group around Wolchok was able to compromise this system because of a simple
bug in the code, which was supposed to read the PDF file used as a ballot~\cite{Wolchok2010}.

\subsection{Summary}
Therefore, open source enables inspecting the code prior to an election to find critical security issues, which
might have been found or exploited by an attacker in the real election. It is also useful to publish the code for
verification and setting up her own voting system to find critical issues and reporting them prior the election. This
will preserve attacks like they were possible in the first phase of iVote or DVBM and should therefore be used for a
voting system.

The philosophy of Civitas is the way to go. All parts of the software are published and documented. The Estonian
system is also a good example, but for the overall transparency they should make it completely open source. Our
suggestion is then to \textbf{open source} the complete application to make it accessible for the community and research
groups.

%%%%%%%%%%%%%%%%%%%%%%%%%%%%%%%%%%%%%%

\section{Anonymous Communication}
\label{anonymous-communication}
The infrastructure and the general design philosophy are now defined. As a next step, we need to think about how to
connect to the voting system, without risking that an attacker might listen to or manipulate the communication between
the voter and the election's servers.

When designing an electronic voting system, we must provide an untappable channel for the complete communication between
voter and server to prevent attackers to observe the traffic (see~\ref{estonia-security-problems}). This channel must be
indistinctive from the normal traffic. Otherwise it is trivial for an observer to verify that the observed sends
requests to the voting server, when only these requests are exclusively sent through this anonymous channel.

\paragraph{Low-Latency-Networks and Timing Attacks}
Some voting systems suggest Tor\footnote{\url{https://www.torproject.org}} or similar software for the untappable
channel. But for network attackers it is easy to run timing attacks~\cite{EnglishErinandHamilton1996} and analyze this
traffic to see if a voter voted or not and low-latency networks like Tor are no protection against it. In our
paper~\cite{TorIsNotEnough2015} we were able to show that it is already sufficient to observe the traffic of the ballot
boxes and some coerced persons to make it very likely to prove, whether the voter placed her ballot and stayed away of
the election or not. Assumed we know the typical traffic-pattern when placing a ballot, then this timing attack is
enabled. We discuss this in subsection~\ref{coercion-open-issue}. It is a non-deniable possibility to coerce a complete
group of voters.\\
If Tor or a similar low-latency network is exclusively used to cast a vote, it is trivial to see if an observed user
connected to the voting server, because this will be the only traffic passing an anonymous network.

\paragraph{High-Latency-Networks}
On the other hand, there exist high-latency-networks, where packets travel long time from client to server. This will
eliminate timing attacks, but is unpractical for an electronic voting system, because the user directly needs feedback
if her vote was counted and not some minutes later. Placing the vote without getting feedback even might decrease trust
into the system. Therefore, the untappable channel should not be implemented through a system with too high latency as
it is common in high-latency-networks.

\subsubsection{Current Approaches}
Current voting systems do mostly not provide an own untappable channel. Some theoretical systems try to build their own
mix-nets, but since coercion starts with the connection to the server (the voter might not have placed her ballot up to
this point), this is not a sufficient protection against network-level attackers. A normal user does not use anonymous
networks, whereby it would be suspicious and detectable if she uses a network like Tor just for the voting process.

It is an \textbf{open question} how to hide from network-level attackers and is not in the scope of this thesis.
Currently, there are no approaches to solve this issue and since low-latency and high-latency networks alike seem not
applicable it is questionable, whether there is a solution at all.

%%%%%%%%%%%%%%%%%%%%%%%%%%%%%%%%%%%%%%

\section{Ballots}
We described the registration, general architecture and the communication between voter and server. Now it is important
to focus on the ballots themselves, since they are essential for an election.

\subsection{Composition}
\label{composition-of-ballots}
Ballots can be encrypted and have different data structures or properties based on the chosen cryptographic scheme. We
will now present our ideas for the ballot's composition.

\subsubsection{Is encryption really needed for the ballots?}
To put it briefly: yes. Encryption is essential, since it keeps the ballot's content private even if an attacker has
access to the encrypted ballot; she is still not able to see for whom the voter made her sign. This ensures privacy of
the voter's choice.\\
Additionally, the voting system should require a signature by the voter to ensure the ballot's integrity. It is signed
with the voter's private key from her personal eID card, which can be later used for verification.\\
Also a timestamp should be added to the signature for internal use to make it possible to sort the ballots and to
discard old ballots.

Ballots must be secured and safely stored. If the voter does not trust in it, the system will not have any chances to be
accepted. Missing trust in the system is one reason why the Norwegian system was shut down and the trial of internet
voting was cancelled in 2014~\cite{NorwayEndOfficial2014}. This underlines the importance of this point.

During the design process of a voting system, we have to choose which encryption to use. Due to the general architecture
of the problem, it seems natural to use some kind of asymmetric encryption, because we want to encrypt the ballot in a
way, where only one specific authority is able to decrypt it. Now, if we think one step further, we have to keep in
mind that we need to reliably anonymize the ballots before the tallying process, as it is described in
subsection~\ref{anonymization-of-ballots}.

\subsubsection{Homomorphic Encryption}
To support mix-nets, it is necessary to choose a cryptographic primitive, which supports re-encryption. Otherwise
mix-nets make no sense and provide no useful contribution for anonymity (see~\ref{mixtyp}). As we suggest to
use mix-nets for the ballot anonymization (see~\ref{anonymization-of-ballots}), we have to use a homomorphic
encryption scheme which supports re-encryption. So, we suggest to use a \textbf{homomorphic encryption scheme} for the
election's key-pair.

\subsubsection{Threshold Encryption}
\label{construction-threshold}
Some encryption schemes support threshold encryption (see~\ref{diselg}). Distributing the private key, which is needed
to decrypt the ballots, ensures that more than one authority needs to be compromised to decrypt the ciphertexts.
Threshold encryption should be used, since this makes it harder for an attacker to get the ballot's content.
Additionally, corruption is unlikely, because more than just one authority needs to be corrupted to get the key.\\
Since we suggested a \textbf{homomorphic encryption scheme}, it should support \textbf{threshold encryption}, because it
adds more trust into the authorities and makes it much more difficult to corrupt authorities without noticing it. For
maximum security all trustees of the system should have to cooperate to reconstruct the private key, because even one
compromised trustee is enough to cancel the election.

\subsubsection{Summary}
To support threshold- and re-encryption, we suggest to use \textbf{ElGamal} as the chosen algorithm for encryption.
Libraries implementing ElGamal are developed for many programming languages, like Python, where ElGamal is part of the
Python Cryptography Toolkit
\emph{pycrypto}\footnote{\url{https://github.com/dlitz/pycrypto/blob/master/lib/Crypto/PublicKey/ElGamal.py}}.\\
The election has to create an ElGamal key-pair with a sufficient secure key-length, whilst the public key is used by the
voters to encrypt their ballot and the private key is distributed with ElGamal's threshold encryption property. A
sufficient key-length depends on current security standards. The NSA suggests in 2015 to use a minimum key-length of
3072 bits for their TOP SECRET documents~\cite{BlueKrypt2015}, which should also be applicable for the ballots.\\
Additionally, the ballot's integrity is ensured with the voter's signature generated with her eID card. When storing the
ballots, a timestamp should be added to each ballot to support multiple votes (see~\ref{construction-reduce-coercion}),
because this timestamp enables sorting of the ballots to keep only the most recent vote of a voter.\\
The threshold encrypted private key ensures anonymity of the ballots, because the election's private key must be
reconstructed with the election's trustees (see~\ref{diselg}), which is why nobody is able to decrypt the ballots before
they are anonymized.

\subsection{Filtering the Ballots}
\label{filter-ballots}
It is allowed to cast double votes to reduce coercion in this system. So, we have to keep the most recent ballot from
each voter and discard the others. Since each ballot has a signature, we can verify which encrypted ballot belongs to
whom. Then we can sort them by their timestamps and simply keep the latest ballot. The timestamp should only be used for
internal use. Otherwise a coercer could be able to verify if a voter re-voted.

Modified ballots have a wrong signature and are not stored at all, because the verification for the eligible voter would
fail. But votes containing a correct signature with a malicious or simply false input can not be detected in this step,
since the private key is shared and we had to decrypt the ciphertext to reveal the false input. Therefore, this must be
done shortly before the vote tally, where all ballots are decrypted one by one.

\subsection{Anonymization of Ballots}
\label{anonymization-of-ballots}
The assumptions and suggested mechanics from this chapter limit the remaining cryptographic primitives we can choose.
For example, there exists currently no practical usage of blind signatures, when multiple votes are still allowed,
because ballots with blind signatures leave no trace to the voter. Therefore, it is not possible to assign the ballots
to a voter to keep only the most recent vote (see~\ref{blisig}).\\
Since we suggested to enable multiple ballots, we need a different approach for the anonymization of the ballots.

Voting systems which use homomorphic encryption, like the Norwegian or Civitas, rely on mix-nets. This is trustworthy
as long as at least one mix-server is not corrupt. After each step of the mix-net, the ballots are re-encrypted and each
mix-server then publishes a ZKP to verify the mix and the re-encryption. Estonia does not seem to use mix-nets, since
they only describe that they strip off the signature from the ballots and leave a set of anonymous
votes~\cite{Halderman2014}.\\
In DVBM, the authenticated user has to download a PDF file and re-upload it to the web server. There is no description
about the anonymization process, but it is necessary, since the voter adds at least meta information to the PDF file
when she edits and saves it. The maintainers only describe how the ballots are transferred to a non-networked computer,
which tallies the ballots, but there are no information about anonymization techniques.

\subsubsection{Using Mix-Nets}
As suggested in section~\ref{coercion-freeness-constr}, we choose multiple votes as the preferred way, since it
drastically reduces coercion. Therefore, blind signatures can not be used any more, because this is not possible in
combination with the multiple-votes-policy. So, we suggest to use a \textbf{mix-net} as it provides a reliable mechanism
to anonymize the ballots; it is very easy to understand and all steps can be published for universal verifiability. To
fulfill all cryptographic requirements for a mix-net, our homomorphic encryption scheme must support
\textbf{re-encryption}.\\
After the mix-net, all ballots are anonymized and still encrypted, so that even the stored ballots leave no connection to
the voter.\\
Civitas and Norway already describe the correct usage of a mix-net. Therefore, we can follow their
implementations~\cite{CivitasOnline}.

%%%%%%%%%%%%%%%%%%%%%%%%%%%%%%%%%%%%%%

\section{Tallying Process}
\label{construction-tally}
When the voters placed their ballots and the election has ended, the next step will be to collect all votes from the
ballot boxes to tally the anonymized ballots.\\
Depending on the cryptographic primitives the system uses, there are different possibilities to tally the ballots.

\paragraph{Asymmetric Encryption}
The ballots are encrypted with the election's public key. Therefore, each ballot needs to be \textbf{first decrypted and
can than be tallied}. This is a linear operation, because the computational time increases linear with the amount of
ballots.

\paragraph{Mathematical Operations on Ciphertexts}
Norway intends to use homomorphic aggregation of the ballots, but does not use it yet, because the computational time
needed is too high. Basically, homomorphic aggregation is a great approach for voting systems since we can use the
ciphertexts for mathematical computations. For optimization, Gj{\o}steen tries to minimize the ballot's ciphertext to
get a speedup during the tallying process~\cite{Gjosteen2013}. This might be useful, because this could heavily reduce
the computing time needed to apply the homomorphic property, but is currently not recommended due to these reasons.

\paragraph{Tallying Server}
Before the tallying process starts, the ballot boxes have to merge the votes and transfer them to a separate tallying
server. Moreover, all zero-knowledge-proofs and logs must also be transferred to enable verifiability and to check if
any ballots have been dropped or an insider might try to compromise the election. The tallying server must be absolutely
trustworthy, because this server is used to generate the private key with the help of the trustees, to decrypt the
ballots and then tally the result.\\
In Estonia, the tallying server is air-gapped to protect it against network attacks and viruses. The ballots from the
ballot boxes should be burned to a DVD and then inserted into the tallying server, but in fact the administrators do not
follow this rule and use their own private USB sticks~\cite{Halderman2014}. This is a problem as two scientists from
Security Research Labs showed on the BlackHat conference in 2014 how to hack a USB stick and let it inject malicious
code into the operating system~\cite{USBIssue2014}. Therefore, they can not be sure that the USB sticks are not
modified and using DVDs is the more secure way.\\
The tallying server must be secured and the guidelines for security of the developers should be followed to reduce
possible attacks on this essential server.

\subsubsection{Proof-of-Correct-Decryption}
No matter which encryption scheme is used, the correct decryption and tallying should be verifiable. For this case, we
can use a zero-knowledge-proof, which confirms the correct decryption. This proof is called
\emph{proof-of-correct-decryption} and is described in~\cite{Abe2000}. If the ballots are all correctly decrypted and
verifiable, the authorities can start to sum up the ballots. As a last step, the generated proofs should be published
for verification. These proofs guarantee that decrypted ballots show the intended content of the voter.

Since we chose homomorphic encryption in subsection~\ref{composition-of-ballots}, we have the possibilities to
\textbf{decrypt the ballots one by one} and tally the results. The decryption steps should be made verifiable with
ZKPs.

\subsubsection{Ballot Validation}
\label{validate-ballot}
It is also necessary to \textbf{validate the content} of the ballots. Only in the last step, right before the vote
count, it is possible to catch false input. This should be only a few lines of code, since the structure of a ballot is
kept simple.\\
The ballot's content must also be encoded to prevent malicious code probably being executed on the e-voting servers.

%%%%%%%%%%%%%%%%%%%%%%%%%%%%%%%%%%%%%%

\section{Voter Verifiability}
\label{construction-voter-verifiability}
After we described all steps from the registration to publishing the results, it is important for a user to verify each
step of the election, because this ensures trust in the system and guarantees a correct result. First, we start with the
verification of the voter's own vote.

A voter must be able to verify that her vote was properly counted. This is analogous to how it is done with paper
ballots: the voter authenticates, gets an empty ballot, makes her sign and puts it \emph{herself} into the ballot box.\\
Equally, we need a mechanism in the electronic voting system which shows that the ballot was correctly placed into the
digital ballot box. This must be easy enough, so that the typical voter is able to use this feature and verify that her
own vote was counted.

There are different approaches for verification mentioned in chapter~\ref{systems}: Estonia uses a separate smartphone
application, which needs to scan a unique QR code directly displayed after the vote was placed
(see~\ref{estoniavoting}). This requires that the voter has access to a smartphone and that the Voting Application is
already installed, because she might run out of time to verify her vote; she has only 30 minutes to scan the QR code.
It is beneficial that no other credentials are needed and only the digital identity of the Estonian citizen is
sufficient. But a drawback of the temporary verifiability of the vote is that an attacker just waits these 30 minutes
and manipulates the ballot afterwards when it is no longer possible to verify it (see \emph{Ghost Attack} in
subsection~\ref{estonia-client-side-attacks}).\\
Norway has no eID and needs to send credentials via postal mail, which includes a polling card with codes to compare
them with the code she received via SMS right after the ballot was placed (see~\ref{norwayapplication}). Similarly,
iVote provides a receipt with a number on it, with which she can verify her vote via telephone or Internet
(see~\ref{australiaapplication}).\\
These systems rely on the voter's (mobile) phone for verification. We think this might have a positive impact on the
acceptance of the system, because the user can try the new way to vote and verify it with a trusted device, her mobile
phone.

Since we suggested the trusted electronic ID, it is natural to verify the own vote with the eID. This generates a
receipt, but the voter gets a new receipt each time she places her ballot, which makes it impossible for a coercer to
see, if the presented receipt is for the most recent ballot of the voter or for a discarded one. We will discuss this
later in~\ref{eval-voter-verifiability}.

%%%%%%%%%%%%%%%%%%%%%%%%%%%%%%%%%%%%%%

\section{Universal Verifiability and Publishing the Results}
\label{construction-universal}
After the election ended and the tallying process finished, the results and the ZKPs for each step should be published
online. This way, everyone can verify the outcome of the election.

It is sufficient to publish the outcome and all proofs of the tallying process, e.g.\@ the
\textbf{proof-of-correct-shuffling} after each step of the mix-net and the \textbf{proof-of-correct-decryption}, which
verifies that all ballots were correctly counted after the mix-net~\cite{Abe2000}. Therefore, we suggest to
\textbf{publish all proofs} for verification. These proofs and the voter verifiability makes our system an
\textbf{end-to-end verifiable e-voting system} (see~\ref{preliminary-coercion-summary}), because all steps from placing
the ballot up to the vote count are completely verifiable for the voter.

\subsubsection{Nearly no universal verifiability in common systems}
In fact, the commonly used voting systems do hardly provide any options for universal verifiability. Only Civitas
publishes the anonymized ballots and the ZKPs on a bulletin board. So, the outcome of an election for any of the
real-world systems can not be completely verified and the voter has to trust the election's officials that they properly
tallied the ballots.\\
The Norwegian system added several zero-knowledge-proofs in their voting system, but it is not well documented, if they
are published or not. Basically, there are mechanisms to provide ZKPs for the shuffling and re-encryption of the votes,
which is at least a starting point to universal verifiability.

Combining the transparency of Civitas and the good documentations the maintainers of the Estonian voting system are used
to provide, leads to a comprehensible tallying process. Everyone would be able to verify the outcome with the
published proofs, which fulfills universal verifiability. In this step it is important to keep it simple: if the
verification of the election is too complicated, just few people will try to verify the election.

%%%%%%%%%%%%%%%%%%%%%%%%%%%%%%%%%%%%%%

\section{Summary}
To outline the content of this chapter, we can reduce it into one single table:

\begin{table}[H]
    \centering
    \begin{tabularx}{0.95\textwidth}{l l}
        \toprule
        \emph{Suggested System Components} & \\
        \midrule
        Authentication          & eID\\
        Voting Policy           & Multiple votes\\
        Coercion                & Reduced due to multiple votes and eID\\
        App Structure           & Web\\
        Distributed             & Multiple server locations\\
        Development Model       & Completely open source\\
        Anonymous Communication & Not completely solved\\
        Encryption Scheme       & Homomorphic with threshold-encryption\\
        Ballot Anonymity        & Mix-net with re-encryption\\
        Tallying Process        & Decrypt one by one\\
        Voter Verifiability     & Time restricted verification\\
        Universal Verifiability & Publish ZKPs from \ref{construction-universal}\\
        \bottomrule
    \end{tabularx}
    \caption{Suggested Building Blocks of a Voting System}
    \label{table:ownsystem}
\end{table}

These are the building blocks of e-voting systems built with the cryptographic primitives and techniques which are
currently available and have been approved in real elections. As we can see not all problems are sufficiently solved.\\
Also, we did not focus on common security mechanisms, for example intrusion detection systems and firewalls. This leaves
the scope of this thesis, but should not be forgotten when deploying an electronic voting system.

The next step will be to evaluate these components to verify that they fulfill the preliminaries or where are still
problems, which need to be solved. This is described in chapter~\ref{evaluation}.

\chapter{Evaluation}
\label{evaluation}

After the descriptions of some existing systems and the cherry-picking of their building blocks for our own voting
system, we now evaluate the system from chapter~\ref{construction} against the defined preliminaries from
chapter~\ref{preliminaries}. We will also focus on the weak points and analyze them.\\
Since we designed our system with the preliminaries in mind and with the experiences we gained from real-world systems,
it is not surprising that our system fulfills many of the requirements. Although we only have the theoretical construct
of our system, we will describe how we suggest to fulfill the preliminaries. For a qualified analysis, we should
implement our system, but this goes beyond the scope of this master thesis and needs much more time.

%%%%%%%%%%%%%%%%%%%%%%%%%%%%%%%%%%%%%%%%%%%%%%%%%%%%%%%%%%%%%%%%%%%%%%%%%%%%%%%%%%%%%%%%%%%%%%%%%%%%%%%%%%%%%%%%%%%%%%%%

\section{Evaluate Constructed Voting System}
At first we start with the preliminaries. Since they are essentially needed, it is absolutely important to fulfill them.

\subsection{Eligibility}
We assumed that the government controls a public-key-infrastructure and provides electronic ID cards for each citizen
as it is done in Estonia or to some extent also in Germany (see~\ref{registration-authentication}). With these
eIDs it is very easy to determine if a voter is eligible to vote, since this can be verified when she tries to login at
the election's authentication server. Two ways are conceivable: on the one hand, the government could give a copy of the
electoral register to the election's authorities and authenticate the voter. On the other hand, there might be an API to
query the state of the voter's eligibility.\\
Querying each voter is the more secure way, because the operators of the election have then no access to the complete
electoral register. The API just needs to return \emph{true} if she is eligible and else \emph{false}. But this API must
be provided by the government.

As outlined in the construction chapter, the eID provides the best functionality for authentication and verification of
the voter, because the eID is easy to use and provides the key-pair needed to sign the ballot.\\
How the verification is realized depends on the technical possibilities provided by the government.

\subsection{Coercion Freeness}
Coercion is still a big problem. Our constructed system only \emph{reduces} coercion, but is not able to prevent it
completely (see~\ref{coercion-freeness-constr}). This is achieved with the multiple-votes philosophy and the expensive
credentials (see~\ref{registration-authentication}). Vote-buying is unlikely, because expensive credentials can be used
for multiple services and are not exclusively for an election. But timing-attacks are still possible, as we already
described in~\ref{anonymous-communication}.

Therefore, our constructed system is not completely coercion-resistant, but drastically reduces coercion. This is
currently best practice, but needs more research in this field. The importance of coercion-freeness is discussed in
subsection~\ref{coercion-open-issue}.

\subsection{Availability}
Availability can be achieved by replication and distributing each element of the voting system and to provide mechanisms
against DDoS attacks, like crypto-puzzles. In section~\ref{distributed-infrastructure} we suggested two approaches to
guarantee the availability and stability, whilst we focus here only on the distribution of the servers, because the
blockchain approach currently has too many issues (see~\ref{blockchain-summary}).\\
Servers distributed in multiple data centers provide enough capacity to realize the election. Moreover, the redundancy
of the servers lead to a sufficient availability, since an attacker might attack for example a ballot box in data center
A, but there are still other ballot boxes available in data center B. A ballot box might be compromised and drop or
manipulate ballots, but the voter has the possibility to choose a different box accepting her vote.

Our suggestion was to distribute the servers into multiple data centers, which prevents many kinds of attacks and
increases security and availability because of the stated reasons.

\subsection{Ballot Anonymity and Election Secrecy}
We use the homomorphic encryption ElGamal in our constructed system with the properties of re-encryption and
distribution of the private key (see~\ref{composition-of-ballots}). The election's private key is distributed among the
election's authorities (see~\ref{diselg}), and we defined that all trustees are needed to reconstruct the private key.
We assumed in section~\ref{assumptions} that not all authorities are corrupt. Therefore, the
ballot's anonymity is guaranteed, because there are not enough malicious authorities to reconstruct the key and hence
they are not able to decrypt the ballots.

As a consequence, the ballots are only decrypted after the mix-net and before the vote count, because the election's
trustees reconstruct the private key after the ballots are anonymized. This ensures anonymity and secrecy during the
complete election, which is just what we have to provide in a secure voting system according to
chapter~\ref{preliminaries}.

\subsection{Integrity}
Since the ballots are encrypted using a homomorphic scheme, their content is not alterable without the private key. The
voter's signature, which is connected with the vote until the ballots are anonymized in the mix-net, ensures the
integrity of the ballot. Each step of the ballot anonymization is verifiable with the published zero-knowledge-proof. So
we can assume that the ballot's integrity is granted up to the decryption, which is again proven correct with a ZKP
(see~\ref{construction-universal}).

The ballot's signature is added on the voter's machine and not modified during the complete voting process
(see~\ref{composition-of-ballots}). As long as we can rely on the security of the correctly applied
homomorphic cryptography, we can be sure that the ballot has not been altered and the correct choice of the voter
arrives in the vote count.

\subsection{Correctness}
The proof-of-correct-decryption for all ballots gives us the possibility to verify that the ballot's content has not
been modified in the tallying process. We can then publish the results of the ballots, combined with the ZKPs, to
provide the possibility to verify the outcome. If the sum is in order and the decryption is verifiable, the system is
proven correct.

Combined with the logging server described in~\ref{logging}, we can keep track of the ballots to guarantee that none of
them were dropped. So, if all ballots arrive on the tallying servers and are correctly decrypted, calculating the
correct outcome of the election is trivial and verifiable thanks to the public proofs-of-correct-decryption
(see~\ref{construction-universal}).

\subsection{Robustness}
Malicious ballots or ballot's with false content (i.e.\@ content not belonging to an election) can only be detected
during the tallying process, because the ballot's content is only revealed in the tallying process when the ballots are
being decrypted. There can also be only a maximum of one falsy ballot per eligible citizen, since all older ballots are
automatically discarded and just the most recent ballot is stored (see~\ref{filter-ballots}).

The ballot's content is validated directly after the decryption (see~\ref{validate-ballot}). It includes mechanisms to
encode the ballot's content to provide a protection against code intrusion. If it is correctly implemented, the system
will be robust against malicious code in the ballots, because encoded malicious code can not be executed on the e-voting
servers.

\subsection{Fairness}
If partial results were allowed during an election, the voter would be influenced by these results. She might reconsider
her choice due to these results and vote a different way. But the voter should not be manipulated by early results or
polls, rather she should vote for the party she prefers. Even polls during the elections, which try to construct
pre-results, influence the eligible voters who have not voted yet~\cite{Blais2001}.\\
During the European Parliament elections in 2014, the European Commission warned the countries not to publish their
results until all states have tallied their ballots~\cite{Politico2014}. With this announcement they tried to let all
voters place her vote without the influence of real results, which underlines the importance of this preliminary.

Electronic voting systems have to follow common practices in current elections and must not publish early results. Our
system waits for the end of the election, shuffles the ballots and then decrypts them (see~\ref{construction-tally}).
With this design, it is not possible to get partial results, because only the election's trustees are able to
reconstruct the private key to decrypt the ballots (see~\ref{construction-threshold}), which is done when the election
ended. Thus, this preliminary is completely fulfilled.

\subsection{Receipt Freeness and Voter Verifiability}
\label{receipt-freeness-and-voter-verifiability}
In traditional paper elections, for example in Germany, each voter receives her ballot in a polling place, makes her
private choice and puts the paper-sheet into the ballot box. She does this \emph{herself} and is therefore able to
verify that nobody manipulates or watches her choice until she puts it into the ballot box. After this step, she can
observe the vote tally of the ballots boxes in her district and can verify with her own eyes that no votes are dropped
or manipulated.

For electronic voting systems, we have to implement these steps as similar and comprehensibly as it is stated above.
This is why each voter gets at least a temporary receipt, with which the voter can verify that her vote was counted
(see~\ref{construction-voter-verifiability}). This ensures trust in the system and can easily be implemented
by sending the voter a QR code, a numerical PIN or similar.

This fulfills voter verifiability and receipt freeness, but has some problems, which are discussed
in~\ref{eval-voter-verifiability}.

\subsection{Universal Verifiability}
As seen in \ref{receipt-freeness-and-voter-verifiability}, voters have in classical elections the possibility to observe
the vote count in their electoral district, but not the complete election. With electronic voting systems, it is
possible to make the complete election verifiable with the help of cryptography.

To achieve universal verifiability, it is sufficient to publish the zero-knowledge-proofs after each step of the mix-net
and after each decrypted ballot as described in~\ref{construction-universal}. For this purpose, our system generates the
needed ZKPs for universal verifiability.

It might be possible for an attacker to manipulate masses of ballots in an electronic voting system; only one
vulnerability in the system might be sufficient to compromise (at least one of) the servers to be able to manipulate the
ballots. Since we can never be sure that our system is 100 percent secure, we can provide mechanisms to observe the
election. For this purpose, we suggest the logging server and the zero-knowledge-proofs; they provide verifiable
information, which might reveal the manipulations by an attacker. Thus, universal verifiability is needed and must be
implemented as described in our constructed voting system. The ZKPs, which should be published, are described in
(see~\ref{construction-universal}) and fulfill this preliminary, because the all critical parts of the voting systems,
like the anonymization and decryption of the ballots, provide these zero-knowledge-proofs.

\subsection{Summary}
The previous subsections summarize, which preliminaries are fulfilled and which are not. This table shows a brief
overview:
\begin{table}[H]
    \centering
    \begin{tabularx}{0.95\textwidth}{l l}
        \toprule
        \multicolumn{2}{l}{\emph{Evaluate the Voting System against the Preliminaries}}\\
        \midrule
        Eligibility             & Completely\\
        Coercion                & Reduced   \\
        Availability            & Completely\\
        Anonymity               & Completely\\
        Integrity               & Completely\\
        Correctness             & Completely\\
        Robustness              & Completely\\
        Fairness                & Completely\\
        Receipt Freeness        & Questionable\\
        Voter Verifiability     & Completely\\
        Universal Verifiability & Completely\\
        \bottomrule
    \end{tabularx}
    \caption{Fulfilled Preliminaries of our Constructed Voting System}
    \label{table:ownsystemevaluation}
\end{table}
Some points cannot be completely fulfilled. We will discuss them in section~\ref{open-issues}.

\section{Comparison}
Now, we want to focus on the real-world systems and our own system and evaluate, which of the systems fulfills (most of)
the requirements from the preliminaries. We still have to keep in mind that our system from chapter~\ref{construction}
is still not implemented and exists only as a theoretical construct.

We will use some color codings in table~\ref{table:completepreliminaryevaluation} to visualize whether the system
fulfills the requirements from chapter~\ref{preliminaries} completely, partly or not:
\begin{itemize}
    \item[]
        \begin{tabular}{ll}
            \green  & Preliminary \textbf{completely} fulfilled
        \end{tabular}
    \item[]
        \begin{tabular}{ll}
            \orange & Preliminary \textbf{partly} fulfilled
        \end{tabular}
    \item[]
        \begin{tabular}{ll}
            \red    & Preliminary \textbf{not} fulfilled
        \end{tabular}
\end{itemize}
When it is unknown if the system fulfills a preliminary at all, the cell is left in white color.

Although the receipt-freeness has some problems we are discussing in~\ref{discussing-coercion-issue}, we assume it as
fulfilled, when re-voting is allowed or there is no receipt generated at all.

Applied on the definitions of the real-world voting systems produces a short overview comparing the requirements:

\begin{table}[H]
    \centering
    \begin{tabularx}{\textwidth}{l|Y|Y|Y|Y|Y|Y}
        \toprule
                            & Estonia  &   DVBM   &  Norway  &  iVote   & Civitas  & \emph{Our System} \\
        \midrule
        Eligibility         & \green   & \orange  & \green   & \orange  & \orange  & \green  \\
        \hline
        Coercion            & \orange  & \red     & \orange  & \red     & \orange  & \orange \\
        \hline
        Availability        & \orange  & \orange  & \orange  & \orange  & \green   & \green  \\
        \hline
        Anonymity           & \orange  & \unknown & \green   & \unknown & \green   & \green  \\
        \hline
        Integrity           & \green   & \green   & \green   & \green   & \green   & \green  \\
        \hline
        Correctness         & \green   & \green   & \green   & \green   & \green   & \green  \\
        \hline
        Robustness          & \green   & \green   & \green   & \green   & \green   & \green  \\
        \hline
        Fairness            & \green   & \green   & \green   & \green   & \green   & \green  \\
        \hline
        Receipt Freeness    & \green   & \green   & \green   & \green   & \green   & \green  \\
        \hline
        Voter Verifiability & \green   & \red     & \green   & \green   & \green   & \green  \\
        \hline
        Univ. Verifiability & \red     & \red     & \orange  & \red     & \green   & \green  \\
        \bottomrule
    \end{tabularx}
    \caption{Overview about the Fulfillments of the Preliminaries}
    \label{table:completepreliminaryevaluation}
\end{table}

Compared to Civitas it seems that we only achieve better results in \emph{eligibility} (and coupled thereto
\emph{authorization}). But our suggestions rely on experiences with real-world systems and are more usable than the
complicated implementations provided by Civitas (see~\ref{future-work-usability}).\\
The building blocks of our system and those from Civitas fulfill most of the preliminaries and none of the real-world
systems come even close to the theoretical systems. There are also nearly no possibilities to verify the outcome of the
election, although calculating and publishing zero-knowledge-proofs leaks no information about the ballots. And this
might convince a sceptical voter that she can trust in the election and reassured vote electronically in the next
election.

Some real-world systems could easily implement mechanisms to fulfill the remaining preliminaries (except coercion). For
example the Estonian e-voting system, where availability, anonymity and universal verifiability are not (sufficiently)
fulfilled. Using a mix-net instead of only stripping off the signature and distributing the system into multiple data
centers would provide a verifiable anonymization and a much better availability. If they would generate ZKPs when
tallying the ballots and publish them together with the ZKPs from the mix-net, it would improve the Estonian system to
fulfill as many requirements as our construction. This requires the Estonian e-voting system to choose ElGamal as their
encryption algorithm, because a mix-net needs the re-encryption property. Using mix-nets and publishing all ZKPs would
also improve the Norwegian e-voting system to fulfill nearly all preliminaries.\\
Other systems, like DVBM, are not as easy to improve as the Estonian or Norwegian system, because the general structure
has more flaws, for example the PDF ballots and non-existent client-side encryption of the ballots.

\subsubsection{Influence of one Company -- Scytl}
\label{scytl-company}
It is very interesting to see that the Australian iVote performs much more badly than the Norwegian system. Both
systems are developed by the company \emph{Scytl}, but it does not seem that they used their experiences from the
implementation of the Norwegian system to build iVote; coercion, anonymity, universal verifiability and the general
documentation is much worse than in Norway.\\
We tried to contact Scytl to get some more information about iVote and their implementation, which might show up the
difference, but we only received a white paper defining the company's philosophy for an end-to-end verifiable voting
system~\cite{Khaki2014}, which contains no concrete answers to our questions.\\
Scytl is also the developer of the new Swiss e-voting system, which will release in the near
future~\cite{VotingNews2015}. We are interested in the implementation of this system, because it currently seems that
Scytl reduces security and leaves out other preliminaries for economical reasons. And that is definitely not they way it
should be, when a company with sufficient knowledge and experience in building end-to-end verifiable and secure voting
systems leaves out critical parts for an electronic voting system.

%%%%%%%%%%%%%%%%%%%%%%%%%%%%%%%%%%%%%%%%%%%%%%%%%%%%%%%%%%%%%%%%%%%%%%%%%%%%%%%%%%%%%%%%%%%%%%%%%%%%%%%%%%%%%%%%%%%%%%%%

\newpage
\section{Open Issues}
\label{open-issues}
Some preliminaries could not be fulfilled or at least need to be discussed further. In this section we will concentrate
on them and also take a closer look on problems of the real-world elections, which are also relevant besides fulfilling
the preliminaries.

\subsection{Coercion}
\label{coercion-open-issue}
Coercion is one of the few points, which can not be completely solved. In the best case we can reduce it, which is not
really a satisfying solution. We now want to present an attack vector, which enables timing attacks and thus coercion.

\subsubsection{Anonymous Communication between Voter and Voting System}
\label{eval-anonymous-connection}
The preliminaries do not elaborate on the anonymous connection between voter and the ballot box. This is an issue,
because timing attacks are possible even when a low-latency network is used. We evaluated this in detail with a toy
example inside a big simulated Tor network in~\cite{TorIsNotEnough2015}. Our strategy was to think about the TCP
transfer pattern when placing a ballot, e.g.\@ a TCP handshake, payload (the ballot) and then closing the connection.
This is a typical pattern, which will look the same with nearly all ballots and enables this timing attack.

\begin{figure*}
    \includegraphics[width=\textwidth]{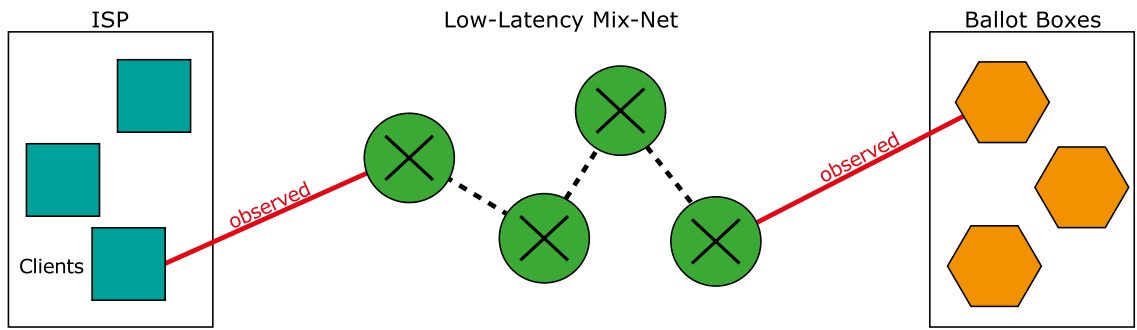}
    \caption{Scheme of Attacker Model \cite{TorIsNotEnough2015}}
    \label{fig:attackermod}
\end{figure*}

In our attacker model, the adversary has access to the traffic of (at least) one ISP and is able to observe the traffic
to (some of) the ballot boxes (see figure~\ref{fig:attackermod}). Now, we can try to find the pattern of sending a
ballot in the TCP dump of the observed traffic. With only few parameters and no special knowledge in finding patterns in
a large amount of data, we were able to find about 95\% of the possible findable voters and could prove, \emph{that}
they placed their vote. Nevertheless, this does \emph{not} mean that we were able to read the content of the ciphertext.

A possible attack on an election might look like this: if an attacker wants to manipulate the outcome of an election,
she could force a whole group of eligible voters to stay completely absent of the election. The network-attacker, who is
able to observe the described parts of the network, has with the TCP pattern-analysis a simple mechanism to prove with a
high percentage, if the voters really stayed absent or not.

This is an attack vector with the possibility to manipulate the results of relevant elections, because there are always
districts known to usually vote for a certain party. Therefore, coercing these regions has a non-negligible impact on
the results.

\paragraph{Discussion}
\label{discussing-coercion-issue}
It is obvious that low-latency networks, like Tor, are susceptible for this kind of attack. A possible solution to
hide the traffic of the ballot, would be noise traffic. But this would also be suspicious and could be recognized by the
attacker, because this is generally just a bigger pattern which must be found.

High-latency networks, like Mixminion \cite{Danezis2003}, Nonesuch \cite{Heydt-Benjamin2006} and Freenet
\cite{Clarke2001}, might be the possible solution, because timing attacks are not possible, but this method does not fit
to electronic voting. How usable is a system, where you have to wait at least four hours until your ballot was counted?
The user has a high interest in verifying her vote \emph{directly} and not much later. Therefore, existing
high-latency networks do not seem to fit with electronic voting.

Placing the ballot boxes inside the low-latency network as a \emph{hidden service} might be a solution, because the
attacker is no longer able to watch the incoming traffic of the ballot boxes. But as long as the user \emph{exclusively}
uses the low-latency network for placing a ballot and normally browses the web without anonymous channel, it is trivial
to find the moment the voter placed her ballot.

We have no complete answer to solve this question. Further research is needed in this regard, because low-latency and
high-latency networks alike do not completely solve coercion or are not suitable for an e-voting system.

This is a big thing, because network-observing attackers are very likely nowadays; the New York Times published a
document from the Signals Intelligence of the United States describing that electronic voting systems even \emph{invite}
them to be exploited~\cite{NYTNSA2013}. From this leaked \emph{SIGINT Mission Strategic Plan FY2008-2013} descends the
quote:

\begin{quote}
    \emph{``Internet-centric activities such as e-commerce, \underline{e-voting}, and on-net industrial and utility
    control \underline{beg to be mined}, even as we expand existing operations against both public and private
    nets.''}
\end{quote}

They did not explicitly say that they observe the network for this kind of attack, but this statement should alarm us
to be prepared for \emph{all} kinds of attacks. Therefore, we can not close our eyes on this possibility to coerce
eligible voters.

\subsection{Voter Verifiability}
\label{eval-voter-verifiability}
Whenever we try to implement voter verifiability, we reveal the ballot's content, because the vote itself or an encoded
version of it is displayed on a screen or similar. It is always possible to make a screenshot or take a photo of it to
get a receipt of the ballot. This breaks receipt-freeness and enables coercion, since the voter is now able to prove
her vote to a coercer.

This receipt is worthless when the voter casts a new ballot. Therefore, she has the possibility to verify her vote,
whereby she again gets some kind of receipt to verify her vote. This is again a (temporary) receipt, which can
be copied and shown to a coercer.\\
With current technologies it is always possible to make a copy of the receipt to get a permanent receipt of the vote.
But if we remove this feature, it would be no longer possible to verify her vote. A system without voter verification is
not end-to-end verifiable and ensures no trust in the users, who are supposed to use the system; they could never be
sure that their vote was really counted properly.\\
A step towards receipt freeness is the way Norway implements voter verifiability: they send a poll card via postal mail
to each voter. The displayed code after the ballot was placed shows the digits of the corresponding party she voted for.
Therefore, the coercer needs a screenshot of the PIN on the display \emph{and} the polling card to have some kind of
receipt. This is a bit more complicated, but has the same problem like normal receipts: she can take a photo of the
polling card with the PIN side by side to have a permanent receipt.

\subsubsection{Other Problems with Temporary Receipts}
A compromised system could easily wait the time a user has to verify her vote, before the attacker changes or drops the
vote. Consider Estonia: the voter has 30 minutes to verify the vote with her smartphone and a QR code. After this
time, the voter has no possibility to do so, therefore she has to trust in the system. But an attacker could drop or
manipulate her vote after these 30 minutes (see~\ref{estonia-security-problems}).

\subsubsection{Negligible due to Multiple Votes}
Since we suggested multiple votes, each receipt is also overridden with a new ballot casted to the ballot boxes.
Therefore, these receipts are not really increasing coercion, because the voter could easily place again her vote, even
after ``proving'' her vote to a coercer; she just casts a new ballot, receives a new receipt and the old one is
replaced.\\
A possible attack vector is still a coercer, who forces the voter to place her vote shortly before the election ends, so
that she has no possibility to place a new ballot, but we do not think that this attack is extensively possible.

It is still questionable whether temporary receipts are conform to the electoral laws of the appropriate country,
because they enable making copies of them to keep a permanent receipt. Obviously, this is no problem in countries like
Estonia or Norway, but this is not assumable in general for all countries.

\subsection{Operational Security and Human Errors}
\label{operational-sec-flaws}
Alex Halderman et al.\@ impressively describe in the papers about the Estonian~\cite{Halderman2014} and the South Wales
voting systems~\cite{Halderman2015} how big the impact of the system administrators are. We already showed a small list
of these problems in the systems in subsections \ref{estonia-security-problems} and \ref{ivotesecurityproblems}.

Most of these problems come apparently because of missing knowledge in security or a general missing sense for it. For
example Estonia: To achieve the biggest possible amount of transparency, the maintainers of the system installed webcams
and streamed meetings to the Internet (without sound, as far as we know), where it was possible to watch an
administrator type in the root password, while the camera was pointing on her keyboard. Or to see the PIN, which is
needed for one of the administrator's eID card or to get access to the wifi-network, because the passphrase was pinned
to a wall.\\
The recordings of these webcams were the basics of the analysis of Halderman et al.\@ to provide founded criticism
against their practices during the election.

A general sense for critical data and security would solve many of the flaws described in these papers. But there must
also be mechanisms to systematically reduce operational errors.

\subsubsection{Bad Implementation}
Errors during the development of the system are somehow natural. But these errors must be reduced, e.g.\@ with hundred
percent test coverage of the code and (peer) reviewing the source code. This reduces unexpected behavior, which was the
entry point for the attack in the Washington, D.C.\@ DVBM voting system: a manually modified version of the
\emph{Paperclip}\footnote{\url{https://github.com/thoughtbot/paperclip}} module was used, which handles file
attachments, like the PDF ballot. Their version was based on an old release of the module and had a shell injection
vulnerability, which has already been fixed in a newer release by the developers of the module and was available when
they deployed the voting system.\\
The maintainers of DVBM were aware of the vulnerability and tried to fix them by themselves. But with no success, which
made the attack of Wolchok et al.\@ possible~\cite{Wolchok2010}.

These are only few examples of problems based on human errors. Many of them can be detected before the official election
when using a complete testing coverage and providing the possibility to get the code reviewed by security experts and
the community.

\subsection{Insider Attacks}
\label{insider-attacks}
An insider is ``\emph{usually a trusted employee, student, or contractor that is granted a higher level of trust than an
outsider}''~\cite{Ruppert2009}. A higher level of trust means in our context that this person has access to (parts
of) the voting system. Among others are those insiders the administrators maintaining the system. In parts of the
webcam-recordings of the Estonian voting systems some of these administrators are logged in as the \emph{root}
user, having total access to the server she is logged on, she could for example copy or drop passing ballots without
leaving a trail to herself.

The ballots might be securely encrypted for nowadays, but there will be computers in the future, which do not need
much time to decrypt the ballots, even without the private key. And an insider, who copied all ballots and took them
with her without being caught, would be able to decrypt the ballots or de-anonymize them. Using everlasting privacy as
described in section~\ref{everlastingprimitive} could decrease the possibility to decrypt the ballots and should
therefore be considered when designing the next electronic voting system.

To prevent insider attacks, a possible solution are users and groups for different levels of permissions. There are
several publications in this field, which can be taken into concern to reduce the possibility of an insider
attack~\cite{Lynch2006, Ruppert2009}.

\chapter{Conclusion}
\label{conclusion}
In this thesis we describe and analyze the general structure of the biggest internet voting systems being used so far.
Many countries are involved in the development of secure electronic voting systems and we could therefore pick the most
relevant around the globe, which were practically used or were supposed to be used in parliamentary elections.\\
Our analysis starts with the description of the \textbf{cryptographic primitives} over the complete voting process up to
the \textbf{security flaws} each system has.

Each analyzed system is unique in the way it solves the problems of electronic voting. All aim to be secure, anonymous
and trustworthy so that the eligible users are ready to participate via Internet in the election. With the knowledge we
gained during the analysis of existing systems and their failures, we \textbf{construct a theoretical end-to-end
verifiable electronic voting system}, which fulfills more preliminaries than the real-world e-voting systems from
chapter~\ref{systems}. Our system also aims to eliminate many of the original system's flaws or at least reduce them.
The construction chapter can be used as a template to implement a voting system, which fulfills most of the
preliminaries.\\
There are still problems with coercion, which need to be solved if the system is supposed to be used in parliamentary
elections. We have no answer how to prevent timing-attacks and are not sure if it is even possible to solve this issue
(see~\ref{discussing-coercion-issue}).

We also focused on a \textbf{peer-to-peer} approach of an electronic voting system. With the help of the
\textbf{blockchain}, we get the possibility to implement a peer-to-peer e-voting system, which does not exclusively rely
on the election's officials and gives each eligible voter the possibility to actively participate in the election.  Some
companies already try to build voting systems with the help of the blockchain, but none of them have produced a working
example.\\
Our contribution to this topic is a \textbf{modified version of the proof-of-stake}, which makes it possible to
construct a secure blockchain without needing masses of computational power to discover the specific hashes needed for
the proof-of-work, which are used by the Bitcoin protocol. This even allows small devices, like smartphones, to
participate in the network to ensure the blockchain's integrity. It can be easily implemented as a small application,
which keeps the voter's device online to work as a full node in the blockchain-based peer-to-peer voting system.

Although we constructed our system with a focus on the security flaws of the real-world systems, we are not able to
eliminate all of these issues, namely coercion. Therefore, we come round to the general opinion of other scientists,
like Alex Halderman, and advice not to use electronic voting systems in coercion-relevant elections, e.g.\@
parliamentary elections, until these open questions are satisfactorily answered. Coercion and other operational flaws
enable the manipulation of an election and must be eliminated, before e-voting should be used in the real-world.

Electronic voting systems should only be used when there are no concerns about their security. These issues might be
solved in the next years and there are many promising publications in this field, but they are temporarily not adaptable
due to their usability.

%%%%%%%%%%%%%%%%%%%%%%%%%%%%%%%%%%%%%%%%%%%%%%%%%%%%%%%%%%%%%%%%%%%%%%%%%%%%%%%%%%%%%%%%%%%%%%%%%%%%%%%%%%%%%%%%%%%%%%%%

\section{Future Work}
We described the basic template with cryptographic primitives and building blocks provided by the best-practices in
network security and with already implemented voting systems. Now, there is still work necessary to a really secure
voting system. Some aspects have already been discussed in sections \ref{anonymous-communication},
\ref{coercion-open-issue}, \ref{eval-voter-verifiability}, \ref{operational-sec-flaws} and \ref{insider-attacks}.

\subsection{Secure Platform}
In section~\ref{assumptions} we assumed that we can trust the voter's computer she is voting with. But many computers
are infected with viruses; some reports say that about 32\% of all computers in the world are infected with viruses and
malware \cite{VirusInfections2013}. The real numbers do not matter, but we can be sure that there will always be
security issues in our computer systems and that there might be some viruses attacking the ballots of electronic voting
systems from the voter's computer. So, we have to research how to prevent malicious software might being able to attack
and manipulate electronic voting systems.

\subsection{Programming Languages and Paradigms}
We criticized in subsection~\ref{jif-programming} that Civitas used Jif in their implementation, which is an unpopular
programming language with only few people being able to use this language.\\
The Cornell University, who invented Civitas, also developed Jif\footnote{\url{https://www.cs.cornell.edu/jif}}, which
extends Java with the support for information flow-control and access control. We should make this an object to analyze
other languages and check, if they are more suitable for security-critical applications, like voting systems.\\
In further research we should also concern about different programming paradigms, like logical or functional
programming, and try to evaluate, which is the best choice in voting systems.

\subsection{Usability Analysis}
\label{future-work-usability}
Several voting schemes have been published, which aim to solve the last issues in electronic voting. These systems are
mostly too complicated or have a very bad usability experience, like Civitas. This is one reason why they are not
successful and are not used for real-world elections.\\
Usability is a very important point: if the voters do not understand how to use the voting system, they will not try it.
Therefore some guidelines should be designed to simplify voting systems and to have a road map, how the components
should look like in the front-end.

\subsection{Implementation}
Theoretical systems have the advantage that they solve all issues in theory, but it is not possible to evaluate them in
real elections. Therefore, this system needs to be implemented for a real comparison with the other big voting systems.

%%%%%%%%%%%%%%%%%%%%%%%%%%%%%%%%%%%%%%%%%%%%%%
%%    End of the main document              %%
%%%%%%%%%%%%%%%%%%%%%%%%%%%%%%%%%%%%%%%%%%%%%%

\backmatter

\bibliographystyle{alphadin}

%\footnotesize{
    \bibliography{library,websources}
%}

\printindex

\chapter*{Ehrenwörtliche Erklärung}

Hiermit versichere ich, die vorliegende Masterarbeit selbstständig verfasst und keine anderen als die angegebenen
Quellen und Hilfsmittel benutzt zu haben.  Alle Stellen, die aus den Quellen entnommen wurden, sind als solche kenntlich
gemacht worden. Diese Arbeit hat in gleicher oder ähnlicher Form noch keiner Prüfungsbehörde vorgelegen.

\vspace{3cm}

\noindent Düsseldorf, 24. September 2015 \hfill Christian Meter

\cleardoublepage

\chapter*{}
\thispagestyle{empty}

\begin{center}
  \vspace{-3cm}
  \fbox{\parbox[c][12cm][c]{12cm}{\centering Hier die H\"ulle\\[1cm]mit der CD/DVD einkleben}}
\end{center}

\vfill

\textbf{Diese CD enthält:}
\begin{itemize}
 \item eine \emph{pdf}-Version der vorliegenden Masterarbeit
 \item die \LaTeX- und Grafik-Quelldateien der vorliegenden Masterarbeit samt aller verwendeten Skripte
 \item die Websites der verwendeten Internetquellen
\end{itemize}

\end{document}